\documentclass[a4paper,11pt]{article}
\usepackage[english]{babel}
\usepackage[utf8]{inputenc}
\usepackage{color}
\usepackage{amsmath,amsfonts}
\usepackage{amssymb}
\usepackage{subcaption}
\usepackage{graphicx}
\color{black}

\newcommand{\st}[1]{\textcolor{black}{#1}}
\newcommand{\redc}[1]{\textcolor{black}{#1}}
\newcommand{\colblue}[1]{\textcolor{black}{#1}}

\begin{document}

\title{Peregrine solitons and resonant radiation in cubic and quadratic media}

\author{%%%% Author details
Marcos Caso-Huerta \\ Department of Information Engineering \\ University of Brescia, Italia \\ Email: marcos.casohuerta@unibs.it \\ {} \\
Lili Bu \\ School of Physics \& Frontiers Science Center\\ for Mobile Information, Communication \& Security \\ Southeast University, Nanjing, China \\ {\&} \\ Department of Information Engineering \\ University of Brescia, Italy \\ Email: lili.bu@unibs.it \\ {} \\
Shihua Chen \\ School of Physics \& Frontiers Science Center\\ for Mobile Information, Communication \& Security \\ Southeast University, Nanjing, China \\ {\&} \\ Purple Mountain Laboratories, Nanjing, China \\ Email: cshua@seu.edu.cn \\ {} \\
Stefano Trillo \\ Department of Engineering \\ University of Ferrara, Italy \\ Email: stefano.trillo@unife.it \\ {} \\
Fabio Baronio \\ Department of Information Engineering \\ University of Brescia, Italy \\ Email: fabio.baronio@unibs.it
}

\date{}

\maketitle
\thispagestyle{empty}\addtocounter{page}{-1} \newpage

\begin{abstract}
We \redc{present the fascinating phenomena} of resonant radiation emitted by \redc{transient} rogue waves in cubic and quadratic nonlinear media, \redc{particularly those shed from } Peregrine solitons, one of the main wavepackets used today to model real-world rogue waves.
In cubic media, \redc{it turns out} that the emission of radiation from a Peregrine soliton \redc{can be attributed to} the presence of higher-order dispersion, \redc{but is} affected by the intrinsic local longitudinal variation of the soliton wavenumber.
In quadratic media, \redc{we reveal} that a two-color Peregrine \redc{rogue} wave can resonantly radiate dispersive waves even in the absence of higher-order dispersion, \redc{subjected} to a phase-matching mechanism that \redc{involves} the second harmonic wave, and \redc{to} a concomitant difference-frequency generation \redc{process}.
In both cubic and quadratic media, we \redc{provide simple analytic} criteria for calculating the radiated frequencies in terms of material parameters, \st{showing} excellent agreement with numerical simulations.
\end{abstract}
%%%%%%%%%%%%%%%%%%%%%%%%%%%

\section{\label{sec:intro}Introduction}
The interest in the generation and \redc{manipulation} of linear dispersive waves has spiked in recent years, \redc{with areas} ranging from optical fibers \cite{Eftekhar2021,Zhang2023}, photonic crystal fibers \cite{Rubino2012,Chemnitz2018}, semiconductor waveguides \cite{Skryabin2017,Rowe2019,Gorbach2010} \redc{to} microresonators \cite{Jang2014,Malaguti2014,Yao2018,Yang2016}, \redc{because of} their critical role in supercontinuum generation\cite{Dudley2006,Skryabin2010} and frequency comb generation\cite{Brasch2016,Okawachi2011,Guo2018}.

\colblue{Starting from the most studied case of solitons arising from the balance of group velocity dispersion (GVD) and cubic (Kerr) nonlinearity, ruled by the nonlinear Schr\"odinger (NLS) equation, it is well known that solitons cannot be in resonance with linear waves since their wavenumbers are intrinsically different. Indeed solitons experience a nonlinear shift of their wavenumber which is of opposite sign compared to the GVD-induced shift of the wavenumber of linear waves. However, the situation changes when the contribution from higher-order dispersion becomes effective, which allows the soliton wavenumber to be equal to that of linear waves with suitable frequency detuning from the soliton central frequency. When such a resonant condition is fulfilled, linear dispersive waves at that frequency detuning can grow spontaneously, by taking energy at the expense of reshaping of the propagating soliton. This phenomenon is usually termed  resonant radiation (RR) \cite{Wai1986,Karpman1993,Afanasjev1996,Roy2009,Milian2014,Baronio2020} or Cherenkov radiation \cite{Akhmediev1995}, and its physical origin is essentially the same (apart from specific details) for solitons which are shape-invariant under unperturbed conditions, higher-order solitons (breathers with zero-background), breathers standing on non-zero background (such as the Peregrine solitons discussed in this paper), or cavity solitons. The net effect is that such wavepackets, which typically show great robustness, give rise to the spontaneous and persistent emission of linear radiation at certain characteristic frequencies detuned from the soliton spectral peak, typically ascribed to the presence of higher-order dispersive effects.
RR has also been found to have played a relevant role in four-wave mixing phenomena \cite{Erkintalo2012,SotoCrespo2012,Conforti2014} and in wave-breaking \cite{Webb2013,Conforti2013b,Conforti2014b}, among others. Furthermore, either built-in or intrinsic periodicity in the system can enhance the emission of RR featuring characteristic multiple spectral peaks under different pumping conditions \cite{Conforti2015,Krupa2016,Nielsen2018}.}

\colblue{We first treat the RR in cubic media \cite{Baronio2020}, using the focusing NLS equation. We will focus on one of the most relevant rational solutions, the Peregrine soliton \cite{Peregrine1983}, which was first observed in nonlinear optics \cite{Kibler2010} but found to be ubiquitous in many branches of physics, including plasma physics\cite{Bailung2011} and hydrodynamics\cite{Xu2019}. Peregrine solitons (oftentimes referred to as Peregrine breathers, as they appear as the infinite period limit for both Akhmediev and Kuznetsov-Ma breathers) are now thought of as the prototype of realistic rogue waves\cite{Solli2007,Dudley2014,Onorato2013}, as they are localized in both space and time and typically present an amplitude more than twice the significant wave height \cite{Kibler2010,Chen2017,Chen2018,Chen2022}.} Its universality relies also on the fact that Peregrine solitons appear as the local waveshape, regardless of the original input shape, in the vicinity of the focusing catastrophe point occurring when nonlinearity dominates over GVD\cite{Tikan2017}. 
%\colblue{For these exotic localized wavepackets, the RR in cubic media is still caused by the higher-order dispersion, but can be affected by the intrinsic local longitudinal variation of the soliton wavenumber, similarly to that shed by higher-order solitons\cite{Driben2015}, but at variance with the radiation shed from ordinary shape-invariant bright or dark solitons.}
\colblue{At variance with the radiation shed from classical shape-invariant bright or dark solitons, but similarly to that shed by higher-order solitons\cite{Driben2015}, for these exotic localized wavepackets the RR in cubic media is caused by the higher-order dispersions and affected by the intrinsic local longitudinal variation of the soliton wavenumber.}
\colblue{The range of physical applications of the RR emitted by Peregrine solitons are very similar to those of classical solitons, but the different conditions for their generation and different regimes in which they operate may make the former RR more suitable for a given setting\cite{Bu2024}, thus considerably broadening the spectrum of applications.}

We \redc{then consider} the RR in quadratic media. \redc{In the past two decades,}  soliton-like pulses in quadratic media have been widely studied in the literature \cite{Buryak2002,DiTrapani1998,Liu1999,Ashihara2002,Moses2006,Marangoni2006,Bache2007}, and the RR \redc{from such solitons} has been both \redc{theoretically}  predicted\cite{Bache2010,Zhou2012,Conforti2013c} and experimentally observed in bulk BBO crystals in the normal GVD regime and effectively defocusing nonlinearity \cite{Zhou2014,Zhou2015,Zhou2015b,Zhou2016}, as well as in periodically poled lithium niobate \cite{Zhou2017}. However, all these analytic predictions have been reliant on the existence of higher-order dispersion. Here we review our recent results on RR \redc{emitted} by Peregrine solitons in quadratic media, both in the cascading  and the phase-matching regimes. Contrary to the cubic case, where RR can only be driven by higher-order dispersion, in quadratic media there \redc{exists another potential mechanism responsible for the generation of linear dispersive waves}. \redc{In latter situation, the RR occurs first at the second harmonic (SH) component, by phase-matching its linear frequency with} the frequency of the Peregrine soliton, and in turn  at the fundamental frequency (FF) \redc{component} via a parametric down-\redc{conversion} process \cite{Bu2022,Bu2023}.

\redc{The subsequent sections are organized as follows. In Sec. \ref{sec:cubic}, we present the Peregrine soliton solution of the cubic NLS equation containing third-order dispersion (TOD) and discuss the RR characteristics caused by TOD. In Sec. \ref{sec:quadratic}, we provide approximate Peregrine soliton solutions in quadratic media and demonstrate their RRs in both the cascading and non-cascading regimes.  Finally, we conclude the review in Sec. \ref{conclu}.}

%%%%%%%%%%%%%%%%%%%%%%%%%%%%%%%%%%%%%%%%%%%%%%%%%%%%%%%%%%

\section{\label{sec:cubic}Cubic media}
We consider the following normalized form of the NLS equation, written in the usual notation for fiber optics:\cite{Baronio2020}
\begin{equation}\label{eq:Cubic}
    i\psi_\xi - \dfrac{\beta_1}{2}\psi_{\tau\tau} - i \dfrac{\sigma_1}{6} \psi_{\tau\tau\tau} + |\psi|^2\psi = 0\,,
\end{equation}
where $\psi$ represents the (complex) wave amplitude, $\tau=(t-z/v_g)/t_0$ and $\xi=z/z_{nl}=z\gamma P$ where $t$ and $z$ are the time and distance in the lab frame, and $t_0= (|\beta''|z_{nl})^{1/2}$ and the nonlinear length $z_{nl}=(\gamma P)^{-1}$ are associated with the power $P$ of the background \st{and the Kerr effective coefficient $\gamma$}; $v_g$ is the group velocity for a carrier frequency $\omega_0$ of the electric field $E(T,Z)=\sqrt{P}\psi(\xi,\tau)$, and $\beta_1=\beta''/|\beta''|$ and $\sigma_1=\beta'''/|\beta''|t_0$ are second and third-order dispersion coefficients, respectively.

\colblue{In the integrable limit $\sigma_1=0$ under the anomalous regime $\beta_1=-1$ (often referred to as focusing NLS), Eq. (1) admits the following rational Peregrine soliton over the unit background:\cite{Peregrine1983}}
\begin{equation}\label{eq:PeregrineCubic}
    \psi(\xi,\tau) = \left[ 1 - \dfrac{4(1+2i\xi)}{1+4\xi^2+4\tau^2} \right] \exp(i\kappa_b \xi)\,,
\end{equation}
where the exponential term comes from the nonlinear Kerr shift of the background. In our normalized setting, the background wavenumber is $\kappa_b=1$ (which corresponds to $\kappa_b z_{nl}^{-1}$ prior to normalization), but henceforth we stick to generic $\kappa_b$ to track its contribution.

This solution behaves as an initially weak pulse over the (unit) background that undergoes a single cycle of compression and growth to finally broaden and decay. This behavior in the time domain is sustained by a strong spectral broadening in the Fourier domain, which serves as a forcing mechanism to generate RR when its frequency is phase-matched via higher-order dispersion.

It also presents a localized shift in the longitudinal phase, which we will denote $\phi_{loc}(\xi,\tau)$ with respect to the background phase $\kappa_b \xi$.\cite{Xu2019} This deviation, which can be seen as a phase anomaly arising from the nonlinear nature of the solution, is intrinsically associated to the localized pulse. Thus, the overall phase of the Peregrine soliton is the sum of the background phase and this local shift, which reads as
\begin{equation}\begin{split}
    \phi_{PS}(\xi,\tau) &= \kappa_b\xi + \phi_{loc}(\xi,\tau) \\
    &= \kappa_b\xi - \tan^{-1}\left( \dfrac{8\xi}{4\xi^2 + 4\tau^2 - 3} \right).
\end{split}\end{equation}
\colblue{Note that, even though the formula for the phase shift appears to present a jump whenever the denominator vanishes, it is just a consequence of absorbing a change of sign in the bracket in \eqref{eq:PeregrineCubic} which does not affect the conclusions, so we will ignore it for the sake of keeping the formulae simpler. The maximum anomaly is attained at the maximum amplitude of the peak ($\tau=0$). The $\xi$-derivative at $\tau=0$ provides the nonlinear wavenumber of the Peregrine soliton:}
\begin{equation}\label{eq:k_PS}\begin{split}
    \kappa_{PS}(\xi) &= \kappa_b + \kappa_{loc}(\xi)\\
    &= \kappa_b + \dfrac{4(8\xi^2 + 6)}{(4\xi^2+1)(4\xi^2+9)}\,,
\end{split}\end{equation}
which similarly consists on the constant wavenumber of the background $\kappa_b$ shifted by a local contribution $\kappa_{loc}(\xi)$. It tends asymptotically to $\kappa_b$ as $\xi\rightarrow \pm\infty$, whereas the local contribution is stronger around $\xi=0$.

A good estimation of the radiated frequencies can be provided by phase matching their wavenumber to that of the Peregrine soliton, $\kappa_{RR}(\omega)=\kappa_{PS}$, where $\kappa_{RR}$ denotes the wavenumber of a linear component wave of the RR at a frequency detuning $\omega$ from the soliton. It is worth noting that for this estimation to be accurate, one must study $\kappa_{RR}$ in a frame making the soliton stationary, hence we need to change to a moving frame with the velocity $v_p=d\tau/d\xi$ of the Peregrine soliton.

Now, to take into account the effects of the full equation (that is, $\sigma_1\neq 0$, making the Peregrine solution an approximate one), we introduce linear waves of the form $\exp(i\kappa_{RR}\xi-i\omega\theta)$, where $\theta=\tau-v_p\xi$ denotes a retarded time in the moving frame, into Eq. \eqref{eq:Cubic}, which, as one can easily check, obey the dispersion relation
\begin{equation}
    \kappa_{RR}(\omega) = \dfrac{\sigma_1}{6}\omega^3 \colblue{- \dfrac{1}{2}}\omega^2 - v_p\omega\,,
\end{equation}

\begin{figure}
    \centering
    \includegraphics[width=0.30\textwidth]{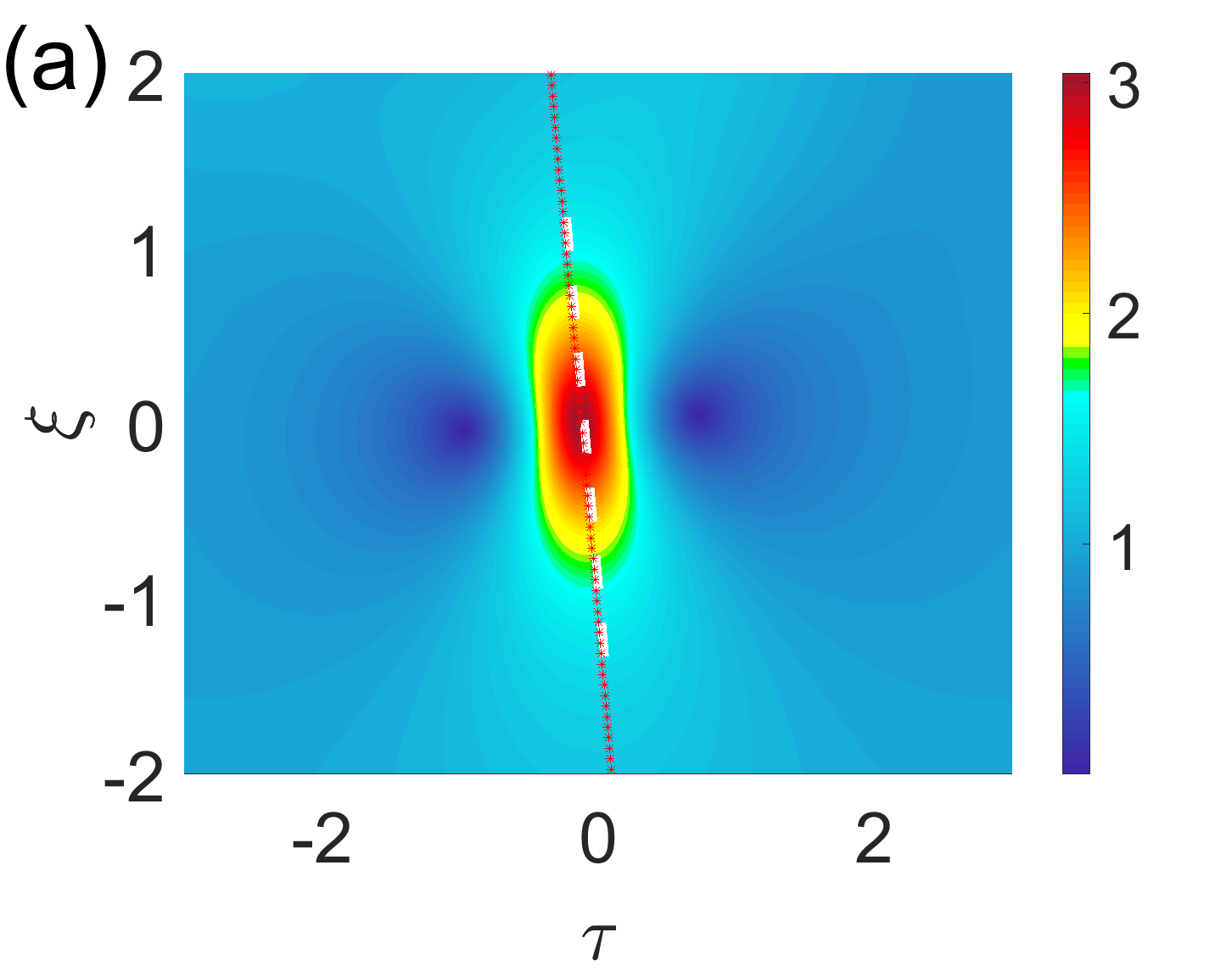}
    \includegraphics[width=0.30\textwidth]{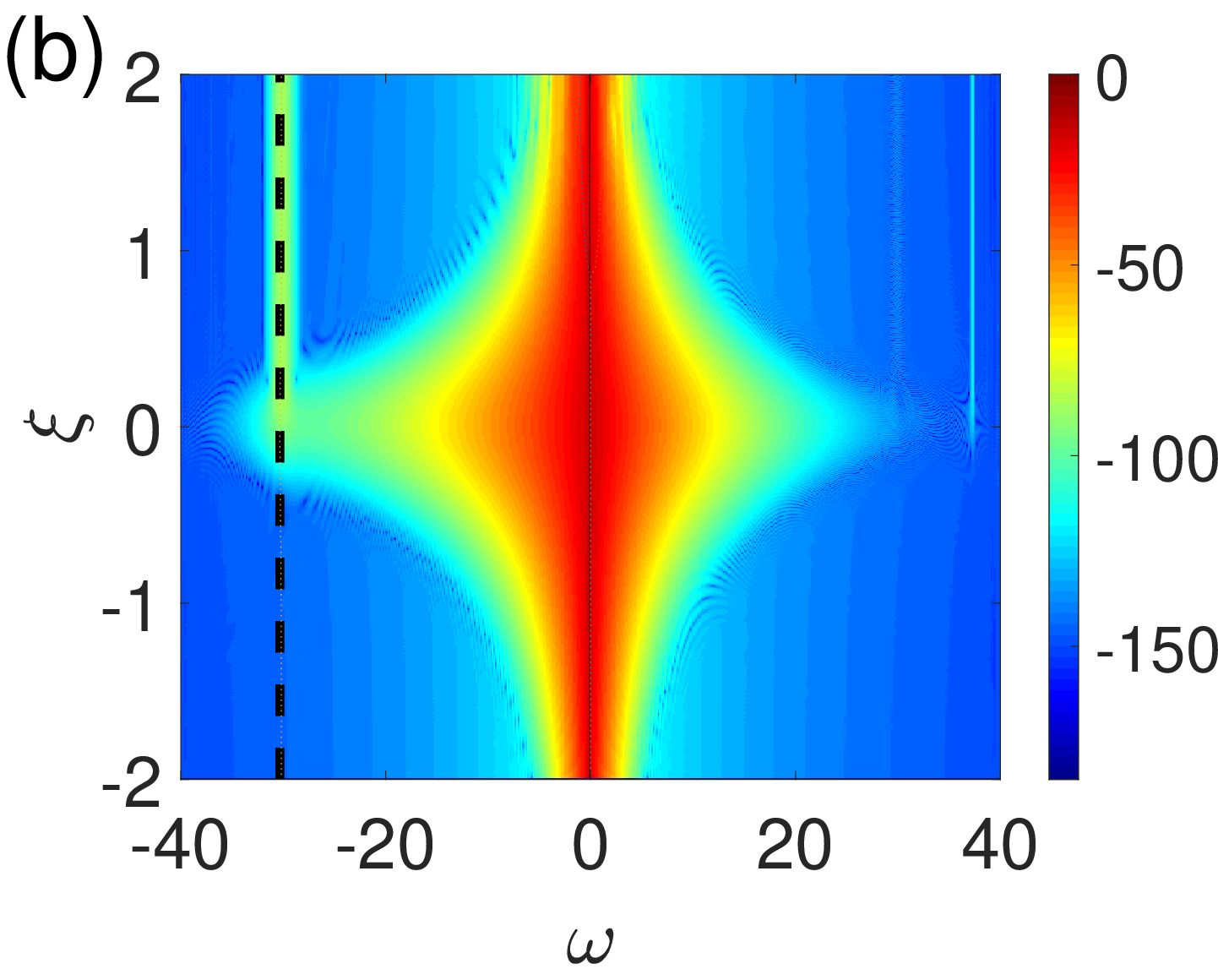}\\
    \includegraphics[width=0.30\textwidth]{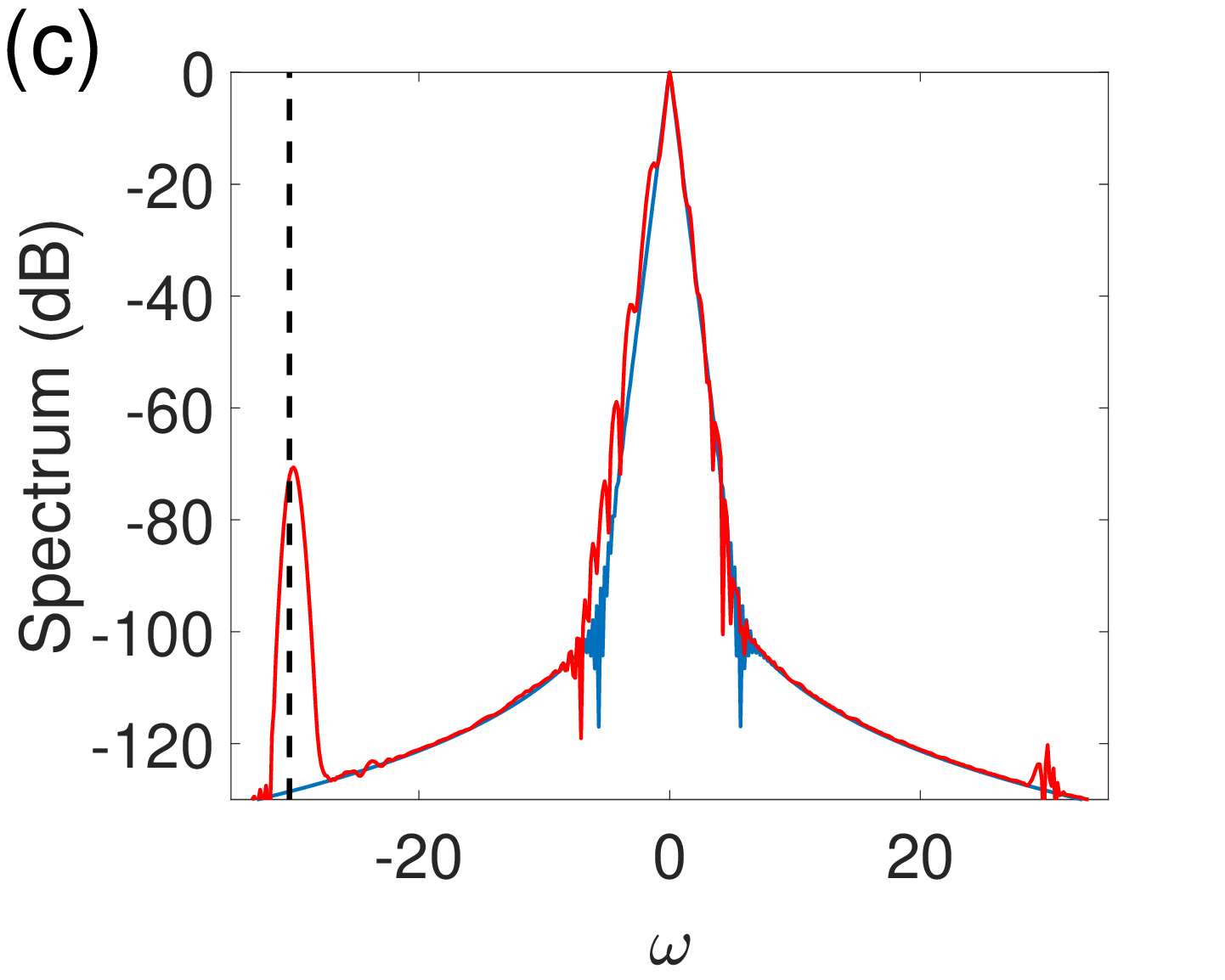}
    \caption{Propagation of a perturbed radiating Peregrine soliton in the cubic case with $\beta_1=-1$ and $\sigma_1=-0.1$: (a) pseudo-color plot of spatio-temporal evolution of the intensity $|\psi|^2$, the dashed white line and dotted red line mark the predicted and numerical Peregrine soliton velocity, respectively; (b) evolution of the Fourier spectrum (in log scale); (c) output spectrum (in log scale, solid red) at $\xi=2$ superposed to the input spectrum (solid \st{blue}). In (b-c) the dashed \st{black} line shows the predicted RR frequency from Eq. \eqref{eq:CubicPM}.}
    \label{fig:cubic1}
\end{figure}

\begin{figure}
    \centering
    \includegraphics[width=0.30\textwidth]{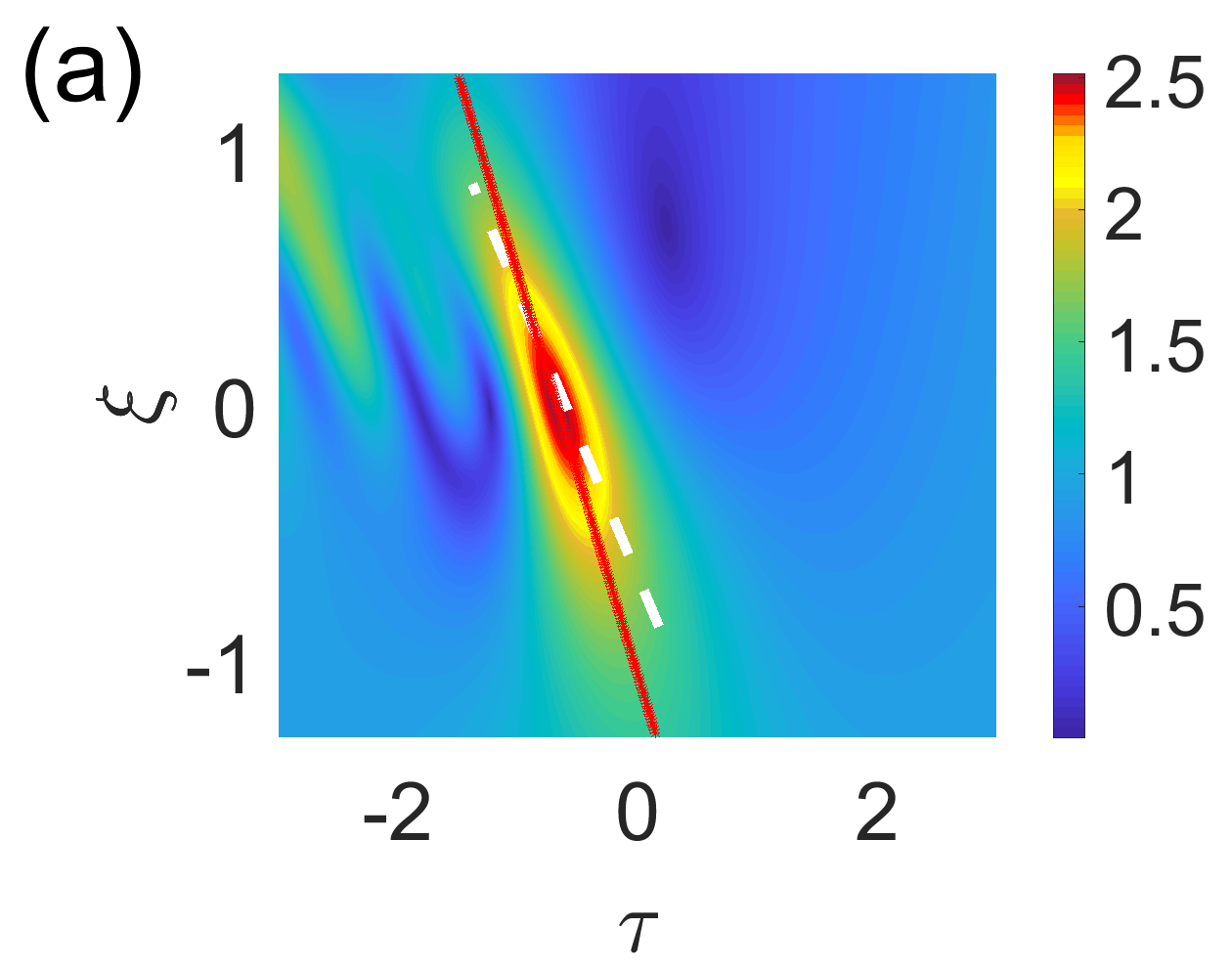}
    \includegraphics[width=0.30\textwidth]{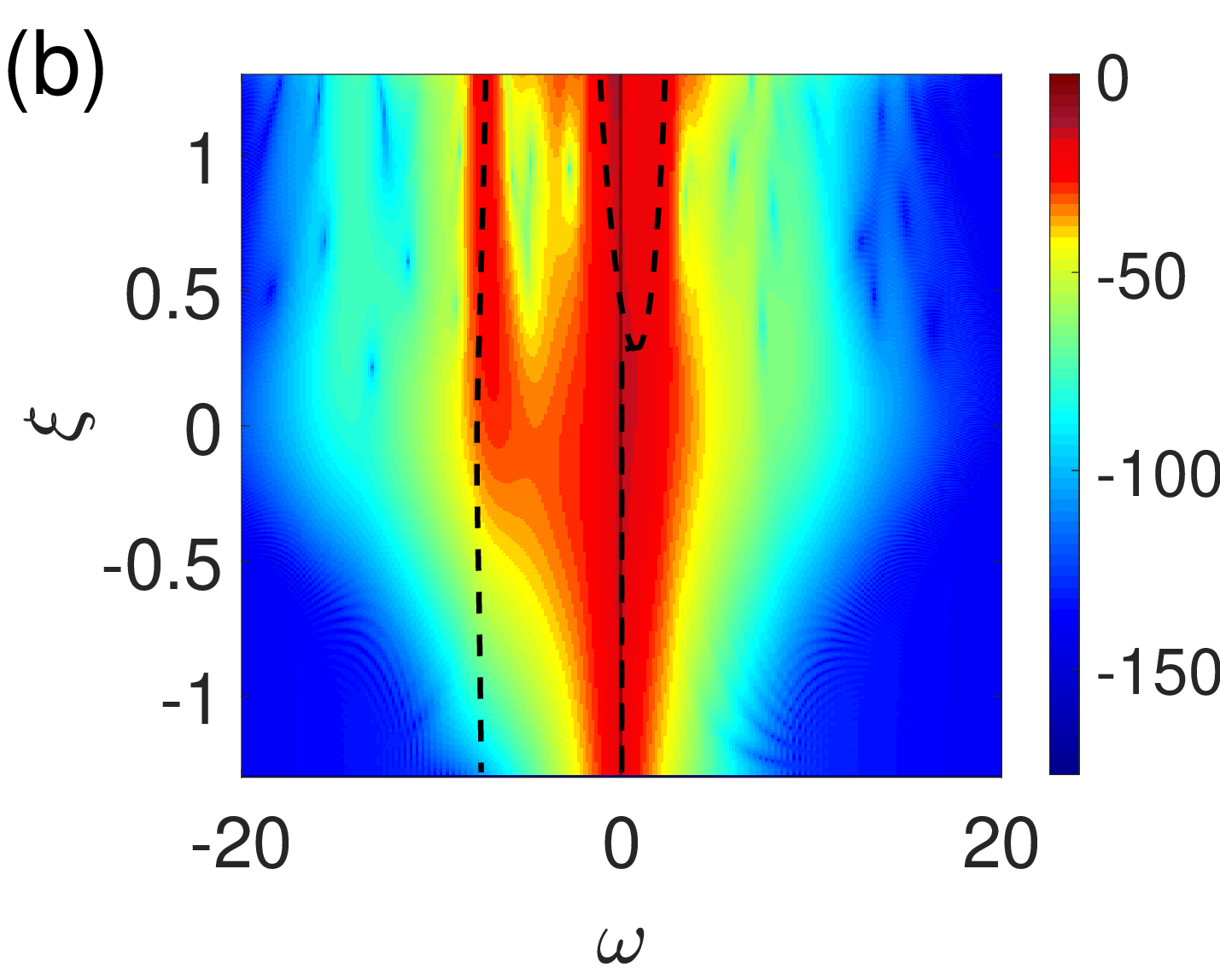}\\
    \includegraphics[width=0.30\textwidth]{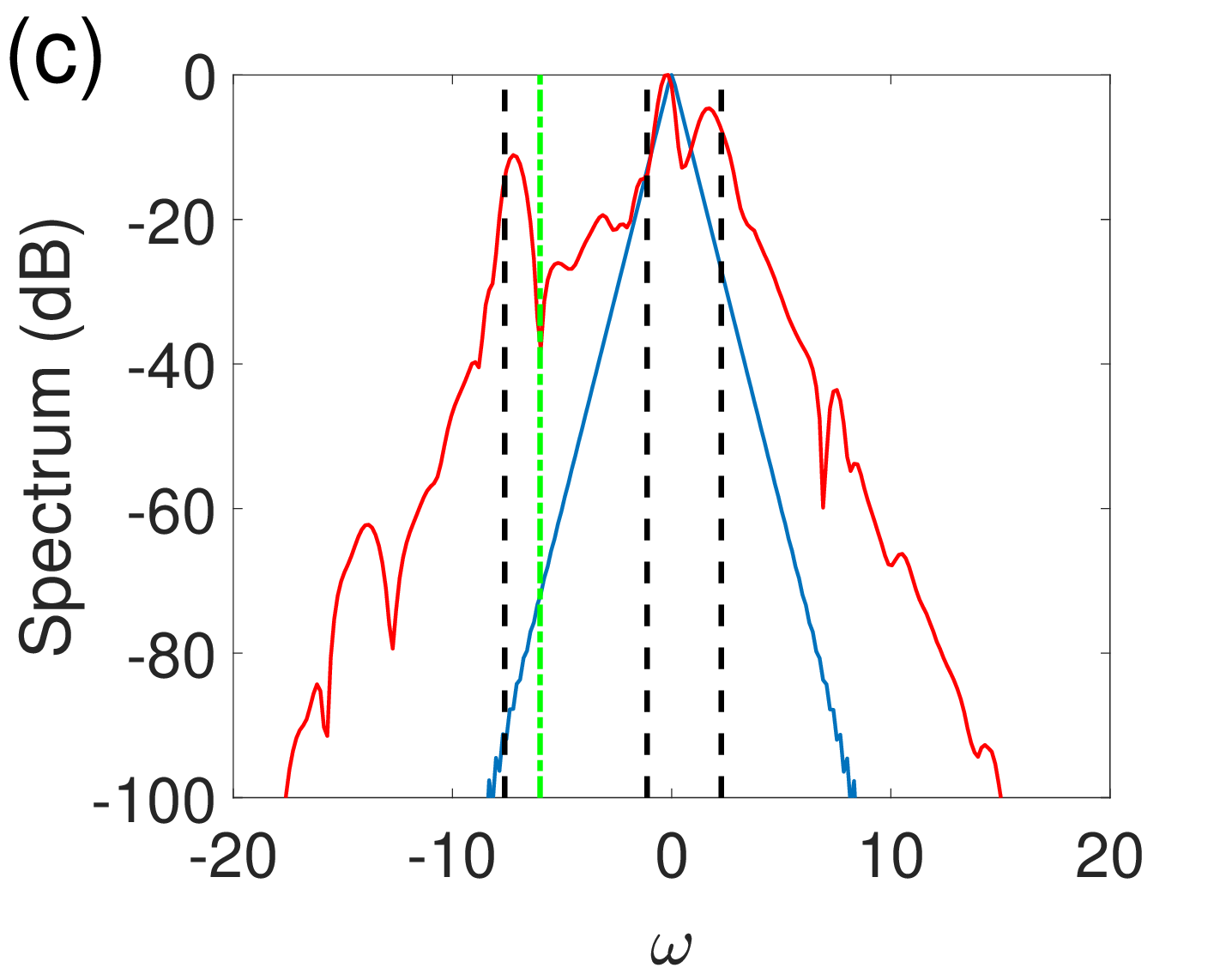}
    \caption{Same as Fig. \ref{fig:cubic1} for stronger TOD $\sigma_1=-0.5$. In (c) the three dashed black lines mark the three predicted RR frequencies, while the dashed green line marks the \st{first-order approximation $\omega_{RR} \simeq \colblue{3/\sigma_1}$.}}
    \label{fig:cubic2}
\end{figure}

\begin{figure}
\centering
\includegraphics[width=0.5\textwidth]{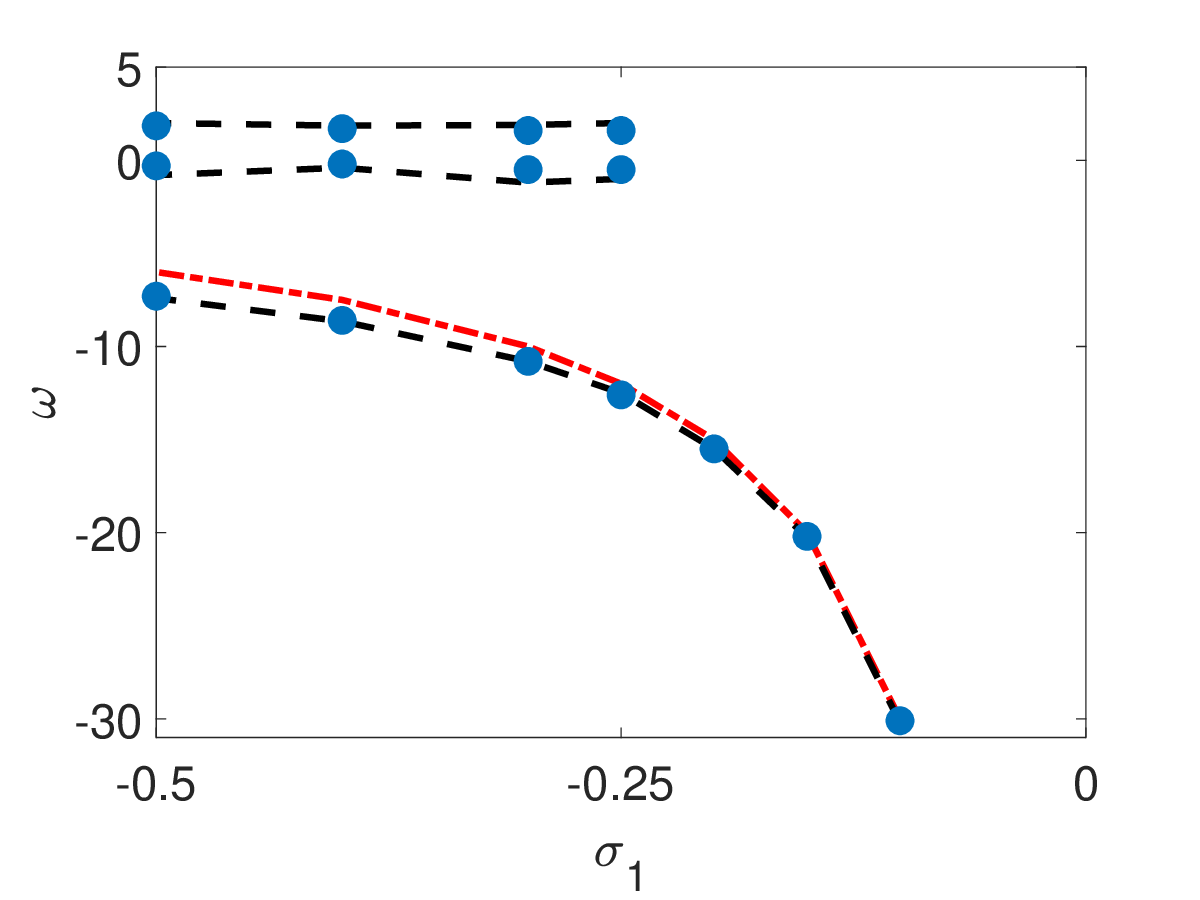}
\caption{\label{fig:cubic3} Impact of the third-order dispersion in the cubic case: frequencies $\omega_{RR1}$, $\omega_{RR2}$ and $\omega_{RR3}$ against the normalized coefficient $\sigma_1$; blue dots: numerical data; black dashed lines: theoretical approximation from Eq. \ref{eq:CubicPM}; red dashed line: first-order approximation $\tilde{\omega}_{RR}=\colblue{3/\sigma_1}$.}
\end{figure}

This expression must however be completed by a nonlinear correction $\kappa_{RR}^{NL}=2\kappa_b$ (which corresponds to $2\kappa_bz_{nl}^{-1}=2\gamma P$ in real-world units) generated by the cross-phase modulation induced by the strong background, which serves as the medium in which the linear waves are generated and propagate. Equating the full form of $\kappa_{RR}$ to the Peregrine wavenumber \eqref{eq:k_PS}, we obtain the phase-matching condition
\begin{equation}\label{eq:CubicPM}
    \dfrac{\sigma_1}{6} \omega^3 \colblue{- \dfrac{1}{2}}\omega^2 - v_p\omega = \Delta \kappa_{nl}\,,
\end{equation}
where the term \st{$\Delta \kappa_{nl}  = \kappa_{PS} - 2\kappa_b = \kappa_{loc}(\xi)- \kappa_b$ }sums up all the effective nonlinear contributions. The roots $\omega=\omega_{RR}$ of Eq. \eqref{eq:CubicPM} provide the resonant frequencies, which correspond in real-world units to $\Omega_{RR}=\omega_{RR}/t_0 = \omega_{RR}(|\beta''|z_{nl})^{-1/2}$.

Note that in order for the standard first-order approximation of the RR frequency, $\omega_{RR}\simeq \colblue{3/\sigma_1}$, with corresponding physical frequency $\Omega_{RR}=3|\beta''|/\beta'''$ (as also estimated from either solitons or wave-breaking \cite{Wai1986,Webb2013,Conforti2013b}), we need both $v_p$ and $\Delta \kappa_{nl}$ to be negligible. In any other case, the phase-matched RR involves the local wavenumber $\kappa_{loc}(\xi)$, \st{unlike the case of both bright and dark solitons}, while the sign of the nonlinear contribution $\kappa_b$ changes as compared to the bright soliton case, \cite{Akhmediev1995}, due to the strong background, similarly to the RR generated by shock waves.\cite{Conforti2014} \colblue{It should also be noted that, in the anomalous dispersion regime the background plane waves are modulationally unstable and may eventually generate additional rogue wave-like peaks due to the introduced perturbations. However, the timespan of these ocurring is much longer than that of the emission of resonant radiation so for our purposes we will treat their effect as being negligible\cite{Newton1987}.}

A typical scenario of RR is shown in Fig. \ref{fig:cubic1}, which depicts the Peregrine solution for $\sigma_1=-0.1$, by launching the exact solution at $\xi=-2$ and integrating numerically Eq. (1) for its evolution. The RR is noticeable in Fig. \ref{fig:cubic1}(a), and even more so in the Fourier spectrum in Fig. \ref{fig:cubic1}(b). Note that the radiating Peregrine soliton drifts slightly with a constant velocity $v_p\neq 0$, which can be estimated by $v_p= \sigma_1 A^2/6$ (plotted in red), where $A$ denotes the peak amplitude of the Peregrine soliton (in this case, $A\sim 3$ for a unit background). The numerical estimation for $v_p$ is also shown in Fig. \ref{fig:cubic1} by a dashed white line. Figure \ref{fig:cubic1} shows that the RR is clearly in the normal dispersion regime and its emission starts when the spectrum of the Peregrine gets broadest, so that it can efficiently seed the frequency $\omega_{RR}$. \colblue{The latter can be estimated analytically as the only real root of Eq. \eqref{eq:CubicPM}. The merit of such estimation is demonstrated in Fig. \ref{fig:cubic1}(\colblue{c}), which superposes the analytical estimation (see dashed vertical line) to the spectrum of the solution at $\xi=2$, \colblue{which features a secondary peak corresponding to the resonant radiation}. A similar analysis for stronger TOD $\sigma_1=-0.5$ (see Fig. \ref{fig:cubic2}), shows that in this case three different frequencies appear (two of which being at much lower detunings), which correspond to the the real roots of Eq. \eqref{eq:CubicPM}. In this case the soliton exhibits a much stronger RR and a much faster drift.}

\st{The effect on the RR frequency of the variations of the TOD coefficient is summarized in Fig. \ref{fig:cubic3}, which report the outcome of different numerical simulations for multiple values of TOD.} As shown, the main branch at large detuning $\omega_{RR1}$ exists for any value of TOD $\sigma_1$, whereas the low-frequency components only start to appear below a certain critical value $\sigma_1\simeq -0.25$, beyond which their frequency is almost invariant to the value of $\sigma_1$. Also note that we have taken the choice of negative $\sigma_1$ but positive values are completely equivalent provided the sign of the RR frequencies is reversed, thus yielding the same physics.

%%%%%%%%%%%%%%%%%%%%%%%%%%%%%%%%%%%%%%%%%%%%%%%%%%%%%%%%%%

\section{\label{sec:quadratic}Quadratic media}
{A different situation occurs when dealing with quadratic media. Contrary to the cubic case, where higher-order dispersion is the driving mechanism that allows for RR to be phase-matched with the soliton, other mechanisms take place in quadratic media. Indeed, when one computes the equivalent of Eq. \eqref{eq:CubicPM} for quadratic media in the cascading regime 
%(i.e., high absolute mismatch yielding an effective Kerr nonlinearity
(where it is possible to reduce the equation to a cubic one) 
\cite{Conforti2013c,Zhou2014}, 
%\colblue{so called because the energy `cascades' from the FF to the SH and then back to the FF, and so forth}) 
without higher-order dispersion, it turns out it has no real roots, thus predicting in principle no RR whatsoever.} 
However, as we will see, a resonance driven by the optical second harmonic (SH) is able to initiate the radiative process which then involves also RR around the fundamental frequency (FF) through non-degenerate down-conversion processes
%thus allowing for RR to take place
\cite{Bu2022}.
Note that extensive numerical and physical experiments have explored the RR in quadratic media but always working under the interpretative hypothesis that the radiation is driven by higher-order dispersive terms
%it is only explored as a consequence of adding higher-order dispersion.
\cite{Buryak2002,DiTrapani1998,Liu1999,Ashihara2002,Bache2010,Zhou2012,Conforti2013c,Zhou2014,Zhou2015,Zhou2015b,Zhou2016,Zhou2017}.

The dimensionless coupled equations for the pulse propagation in dispersive quadratic media are:\cite{Menyuk1994,Buryak2002}
\begin{gather}
    i u_{1\xi} - \dfrac{\beta_1}{2} u_{1\tau\tau} + u_2 u_1^* e^{-i\delta k \xi} = 0\,,\label{eq:Quad1}\\
    \colblue{i u_{2\xi} + i v u_{2\tau} - \dfrac{\beta_2}{2} u_{2\tau\tau} + u_1^2 e^{i\delta k \xi} =0\,,}\label{eq:Quad2}
\end{gather}
where $u_{1,2}(\xi,\tau)$ are the normalized envelopes of the FF and generated SH in the FF group velocity frame, respectively. Here $\xi=z/z_d=z|\beta_1''|/t_0^2$ is the normalized propagation distance, where $z_d=t_0^2/|\beta_1''|$ is the dispersion length. $\tau=(t-z/v_1)/t_0$ is the time in the FF group velocity frame, $v=z_d/z_w$ is the ratio between the dispersion length and the temporal walk-off length $z_w=t_0/(v_2^{-1}-v_1^{-1})$, with $v_{1,2}=dk/d\omega|_{\omega_0,2\omega_0}^{-1}$. $\beta_{1,2}$ are the GVDs, with $\beta_1=\text{sign}(\beta_1'')$, $\beta_2=\beta_2''/|\beta_1''|$ where $\beta_{1,2}''=d^2k/d\omega^2|_{\omega_0,2\omega_0}$. $\delta k=\Delta k z_d$ is the normalized wave number mismatch, where $\Delta k = 2k_1-k_2$ with $k_{1,2}=k|_{\omega_0,2\omega_0}$. Furthermore, $u_{1,2} = \chi z_d A_{1,2}$, where $|A_{1,2}|^2$ measures the intensity, with $\chi=\omega_0[2/(c^3\epsilon_0 n^2_{\omega_0} n_{2\omega_0})]^{1/2}d^{(2)}$ and $d^{(2)}$ is the nonlinear element.

%%%%%%%%%%%%%%%%%%%%%%%%%%
\subsection{Cascading regime}
\colblue{At large mismatches  $|\delta k|\gg 1$ (the so-called cascading regime, where the quadratic nonlinearity mimics the cubic one \cite{Buryak2002}),
by using the SH asymptotic expansion $u_2=\sum_{n=0}^\infty{u_2^{(n)}/\delta k^n}$ and the method of repeated substitution \cite{Menyuk1994}, retaining 
leading-order terms, we arrive at a single FF evolution equation
(obtained at $n=2$ order):
\begin{equation}\label{CLL}
i \rho_ \xi -\frac{\beta_1}{2}\rho_{\tau \tau}+\kappa|\rho|^2 \rho+i \gamma |\rho|^2\rho_{\tau}=0,
\end{equation}
where for simplicity we have defined  $u_1=\rho$, 
$u_2 \simeq \frac{1}{\delta k} \rho^2 e^{i \delta k \xi} + \frac{2 i v}{\delta k^2}  \rho \rho_{\tau} e^{i \delta k \xi}$, 
$\kappa=1/\delta k$, and $\gamma= 2 v/ \delta k^2$.
We emphasize that the mapping of the FF wave evolution of Eqs. (\ref{eq:Quad1}) and (\ref{eq:Quad2}) into 
Eq. (\ref{CLL})  is only valid with large mismatch $\delta k$, and also with large walk-off $v$ compared to dispersions $\beta_{1,2}$. 
Under such conditions, the role of the dispersion $\beta_2$ becomes negligible, and the SH 
wave evolution can be also dictated by Eq. (\ref{CLL}) through the above mapping relation.
}

\colblue{Eq. \eqref{CLL} is known as the integrable Chen-Lee-Liu (CLL) equation\cite{Chen1979},
which when $\gamma=0$ reduces to the integrable NLS equation.}

\colblue{Therefore, we emphasize that approximate two-color quadratic solutions 
$[u_1(\tau,\xi), u_2(\tau,\xi)]$ (so called because of the FF and SH components) of CLL soliton type $\rho(\tau,\xi)$ can be written as:
\begin{equation}\label{map}
\begin{split}
&u_1(\tau,\xi)= \rho(\tau,\xi),   \\
&u_2(\tau,\xi)  =  \frac{1}{\delta k} \rho(\tau,\xi)^2 e^{i \delta k \xi} + \frac{2 i v}{\delta k^2}  \rho(\tau,\xi) \rho(\tau,\xi)_{\tau} e^{i \delta k \xi}.
\end{split}
\end{equation}
}

\colblue{Using the CLL chirped Peregrine solution, obtained via a gauge transformation\cite{Chen2016}, one can write the quadratic Peregrine soliton as}

\colblue{
\begin{gather}
    u_1(\tau,\xi) = \left[ 1 - \dfrac{2i \left( \frac{\gamma\theta}{\beta_1} + \colblue{\kappa\xi} \right) + 1}{\left( \colblue{\kappa} - \frac{\gamma^2}{\beta_1} \right)\left( \colblue{\kappa\xi^2} - \frac{\theta^2}{\beta_1} \right) + \frac14 + i \frac{\gamma(\gamma\xi+\theta)}{\beta_1}} \right] e^{i\kappa \xi}\,,\label{eq:PS1} \\
    u_2(\tau,\xi)=\kappa u_1^2(\tau,\xi) e^{i\delta k \xi}
    + 2 i v \kappa^2 u_1(\tau,\xi)\, u_1(\tau,\xi)_{\tau}\,\, e^{i \delta k \xi},\label{eq:PS2}
\end{gather}}
\colblue{where we have further defined $\theta=\gamma\xi - \tau$, subject to the constraint $\delta k \beta_1 < 0$, necessary for compensating GVD and nonlinearity. Note that the background of $u_1$ was set to have unit value without loss of generality. Different kinds of Peregrine solitons exist for various choices of the mismatch $\delta k$, the GVD $\beta_1$ and the walk-off $v$.\cite{Baronio2017,Baronio2017b}}

%\colblue{It is known that a family of breather-type (i.e., breathing in $\xi$ and periodic in $\tau$) solutions of \eqref{eq:Quad1} and \eqref{eq:Quad2} can be obtained at large mismatches $|\delta k|\gg 1$ (the cascading regime \cite{Buryak2002}). The Peregrine soliton stands for the limit corresponding to a strong double localization in $\xi$ and $\tau$ over a plane wave background.\cite{Baronio2017} }
%\colblue{Here we use the basic two-color solution (so called because of the FF and SH components), presented in Ref. \onlinecite{Baronio2017b}, \colblue{where by using the SH asymptotic 
%expansion $u_2=\sum_{n=0}^\infty{u_2^{(n)}/\delta k^n}$, and the method of repeated substitution \cite{Menyuk1994}, retaining 
%leading-order terms, we arrive at a single evolution equation for the FF (obtained at $n=2$ order):
%\begin{equation}\label{CLL}
%i \rho_ \xi -\frac{\beta_1}{2}\rho_{\tau \tau}+\kappa|\rho|^2 \rho+i \gamma |\rho|^2\rho_{\tau}=0,
%\end{equation}
%where for simplicity we have defined  $u_1=\rho$, 
%$u_2 \simeq \frac{1}{\delta k} \rho^2 e^{i \delta k \xi} + \frac{2 i v}{\delta k^2}  \rho \rho_{\tau} e^{i \delta k \xi}$, 
%$\kappa=1/\delta k$, $\gamma= 2 v/ \delta k^2$. When $\gamma=0$, Eq. \eqref{CLL} reduces to the integrable nonlinear Schr\"odinger equation, whereas when $\gamma\neq 0$ it is known as the integrable Chen-Lee-Liu (CLL) equation\cite{Chen1979}. Using its reported chirped Peregrine solution, obtained via a gauge transformation\cite{Chen2016}, one can go back to our original equation to write the Peregrine soliton as
%}

\colblue{As expected, when reducing it to the focusing NLS equation via the choice of parameters $\gamma=0$ and $\beta_1=-1$, the Peregrine soliton component at the FF reads
\begin{equation}
    u_1(\tau,\xi)= \left[ 1 - \dfrac{4(1+2i\kappa\xi)}{1+4\kappa^2\xi^2+4\kappa\tau^2} \right] e^{i\kappa\xi}\,,
\end{equation}
which coincides with Eq. \eqref{eq:PeregrineCubic} except for the influence of the variable $\kappa$.
}

In general, one of the defining characteristics of solitons in integrable systems is that they do not couple energy into linear waves. However this fact usually does not hold when taking into consideration higher-order terms in the equation which breaks the integrability  (and, in this case, soliton solutions may be only approximate solutions). Based on the fact that the existence of the Peregrine soliton is supported by the interplay between the Kerr-like nonlinearity and the GVD at the FF (where the integrable limit is represented by the NLS equation), we want to seek for RR generated by GVD of the same leading order at the SH in Eqs. (7-8).

RR generated by phase matching of this kind of solution was studied in Ref. \cite{Bu2022}. For the FF component in Eq. \eqref{eq:Quad1}, the condition for the presence of RR was found to be
\begin{equation}
    \label{eq:BreatherCond1}
    \dfrac{\beta_1}{2} \omega_1^2 - \omega_1 v_p = \kappa + \kappa_{loc}\,,
\end{equation}
where $v_p$ is the velocity of the Peregrine soliton, $v_p = 2v/\delta k^2\colblue{(=\gamma)}$, and
\begin{gather}
    \kappa_{loc} = \dfrac{8\kappa \beta_1 - 24\gamma^2}{3\beta_1}\,,
\end{gather}
while for the SH component in Eq. \eqref{eq:Quad2}, the resonant condition was found to read as
\begin{equation}
    \label{eq:BreatherCond2}
    \dfrac{\beta_2}{2} \omega_2^2 - \omega_2 (v_p - v) = \delta k + 2\kappa + 2\kappa_{loc}\,.
\end{equation}
In both cases the method employed to find such conditions was to look for matches between the wavenumber of potentially growing linear wave components of the form $\exp(i\kappa_{\text{lin},\{1,2\}}\xi-i\omega_{1,2}\tau_p)$, where $\tau_p=\tau-v_p\xi$ corresponds to the Peregrine moving frame, with nonlinear wavenumbers $\kappa_{\text{nl}}$ of the Peregrine soliton (i.e., similar to the cubic case \cite{Wai1986,Karpman1993,Akhmediev1995}, though taking into account linear waves around both the FF and the SH). This amounts to neglect nonlinear terms in the interaction between the linear waves and the soliton driving wavepackets. Note that, as compared with Ref. \cite{Bu2022}, we have added the local correction $\kappa_{loc}$ due to nonlinearity for consistency with the derivation for the cubic case. However, as evidenced below by Eq. \eqref{eq:omegawalking} and due to the fact that $|\delta k|\gg |\kappa_{loc}|$ in the cascading regime, its role in this analysis turns out to be negligible.
%and the conclusions with and without it are basically identical.

As expected, the condition on $\omega_1$, Eq. \eqref{eq:BreatherCond1}, \st{does not yield real solutions, consistently with the fact that, in the NLS equation, solitons are not in resonance with linear waves unless higher-order dispersion is present.} Conversely, the condition on $\omega_2$, Eq. \eqref{eq:BreatherCond2}, does yield real roots, which read
\begin{equation}\label{eq:omegawalking}
    \omega_{2,RR}^\pm = \dfrac{(v_p-v)\pm \sqrt{(v_p-v)^2 + 2\beta_2(\delta k+2\kappa+2\kappa_{loc})}}{\beta_2}\,.
\end{equation}

\begin{figure}
    \centering
    \includegraphics[width=0.35\textwidth]{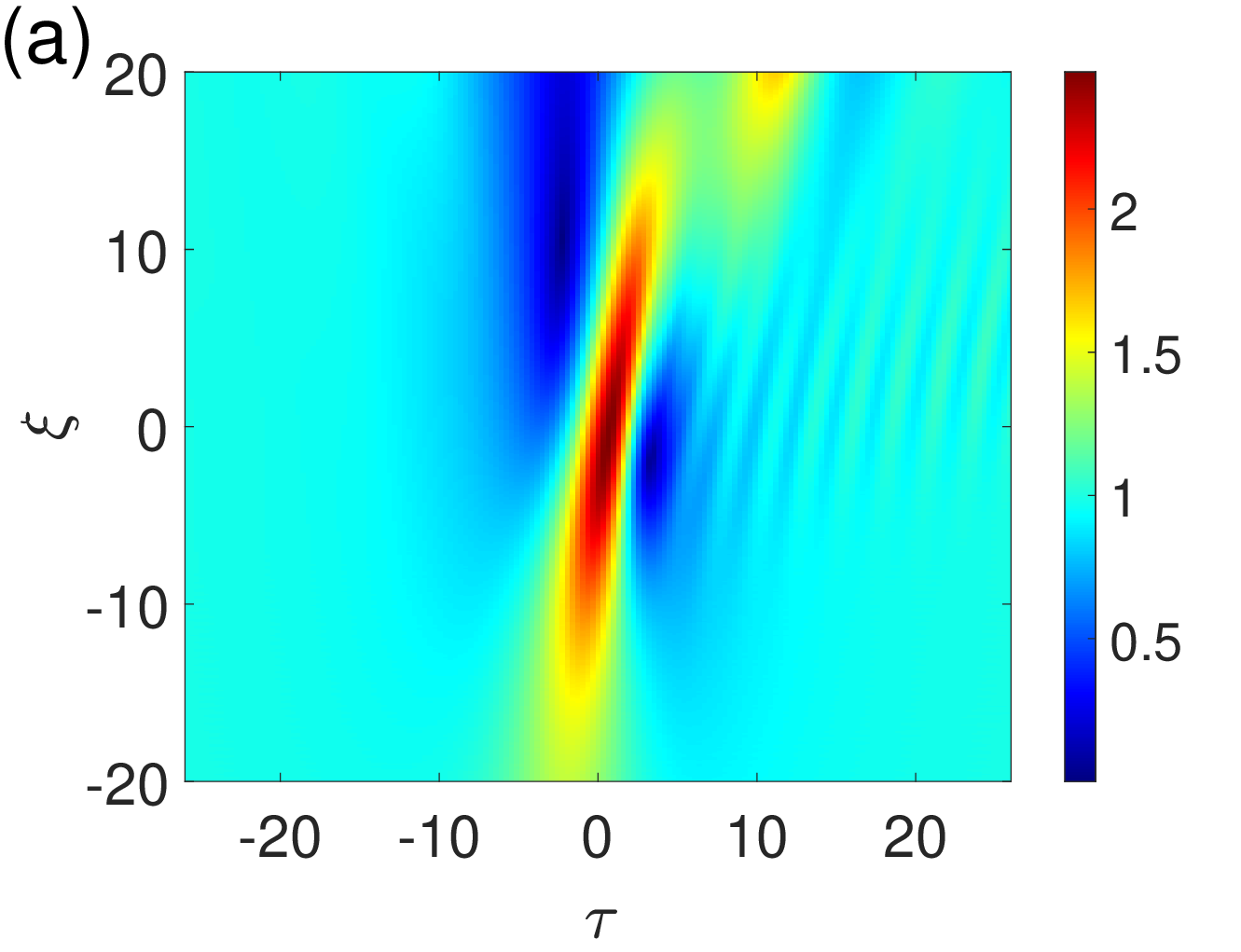}
    \includegraphics[width=0.35\textwidth]{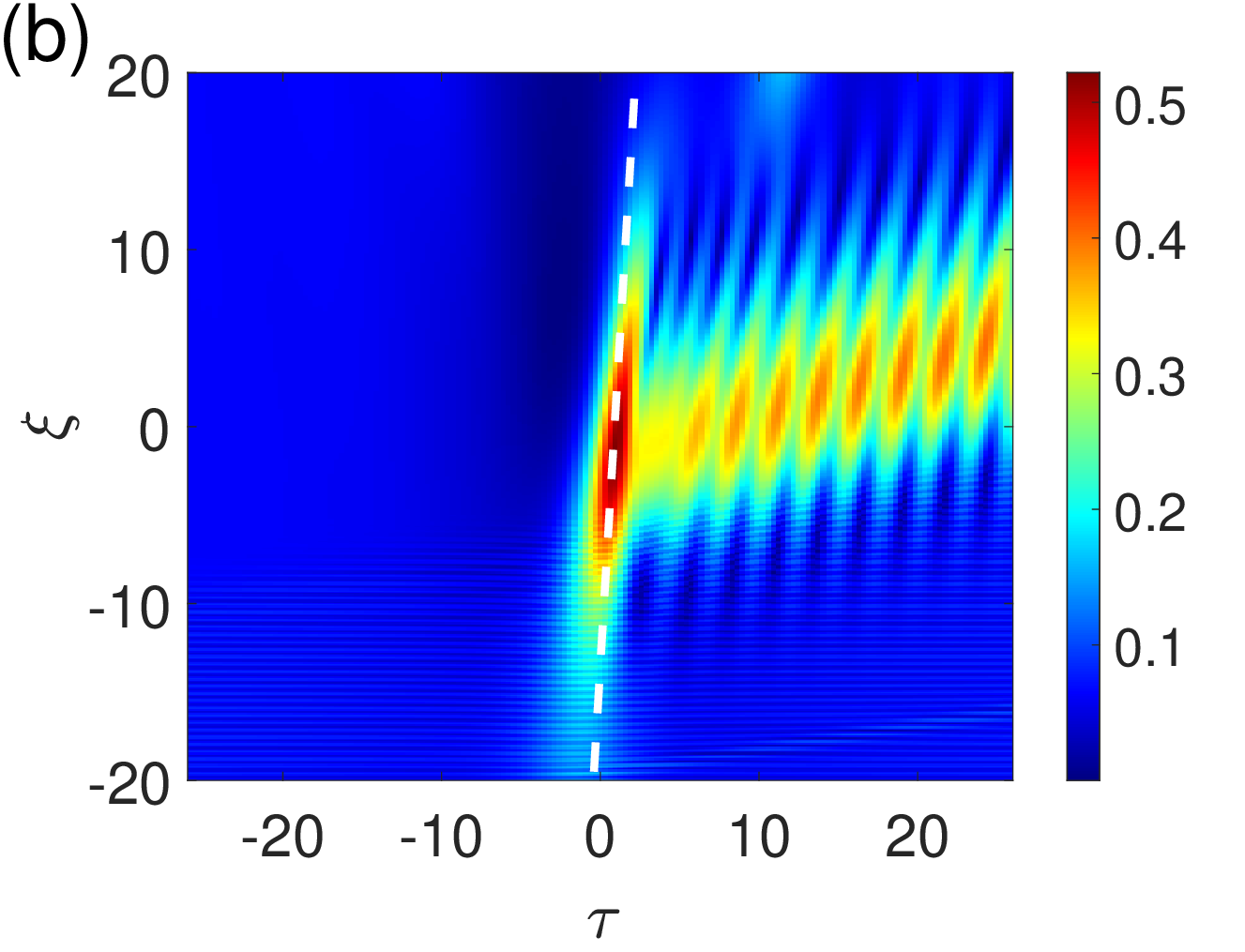}
    \includegraphics[width=0.35\textwidth]{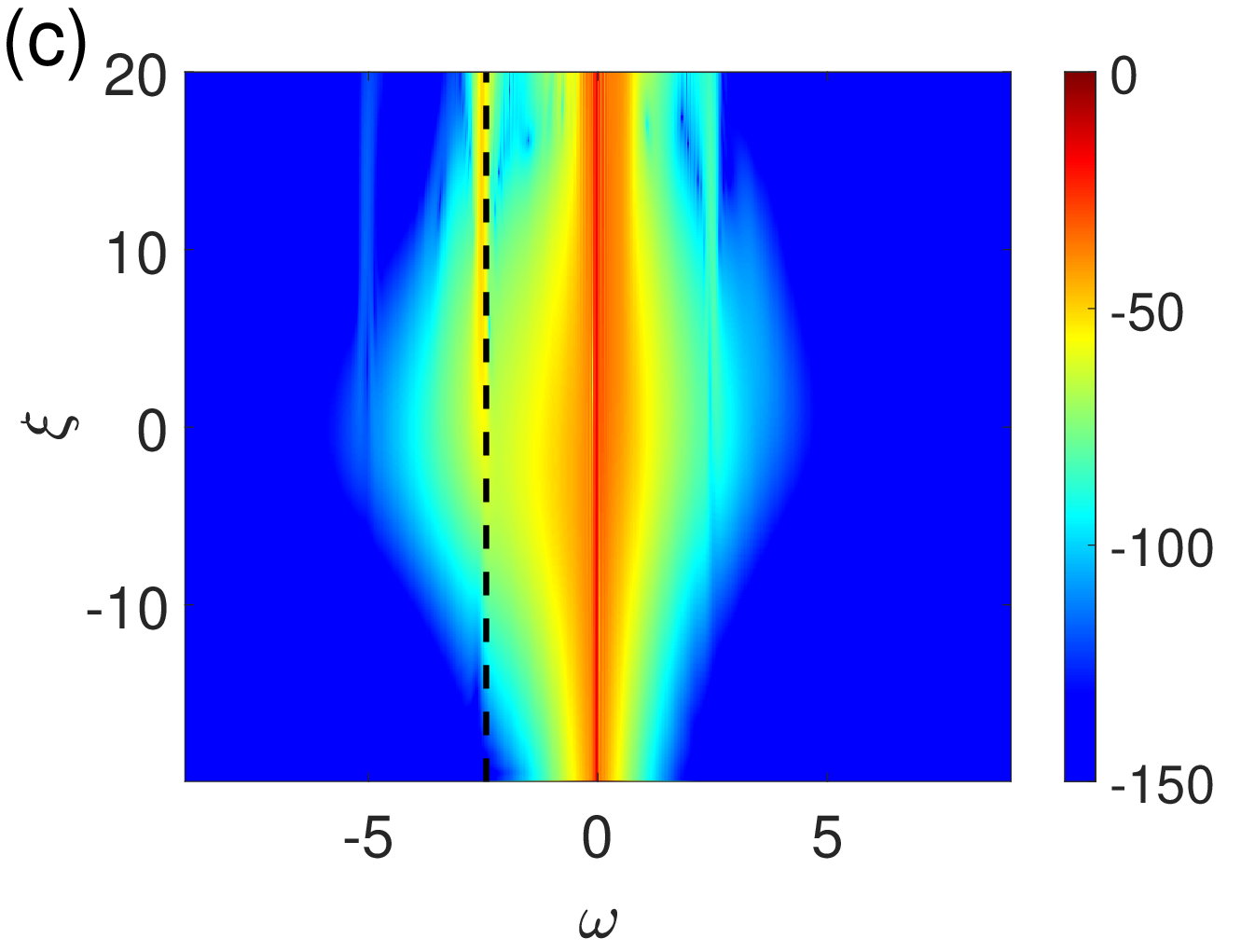}
    \includegraphics[width=0.35\textwidth]{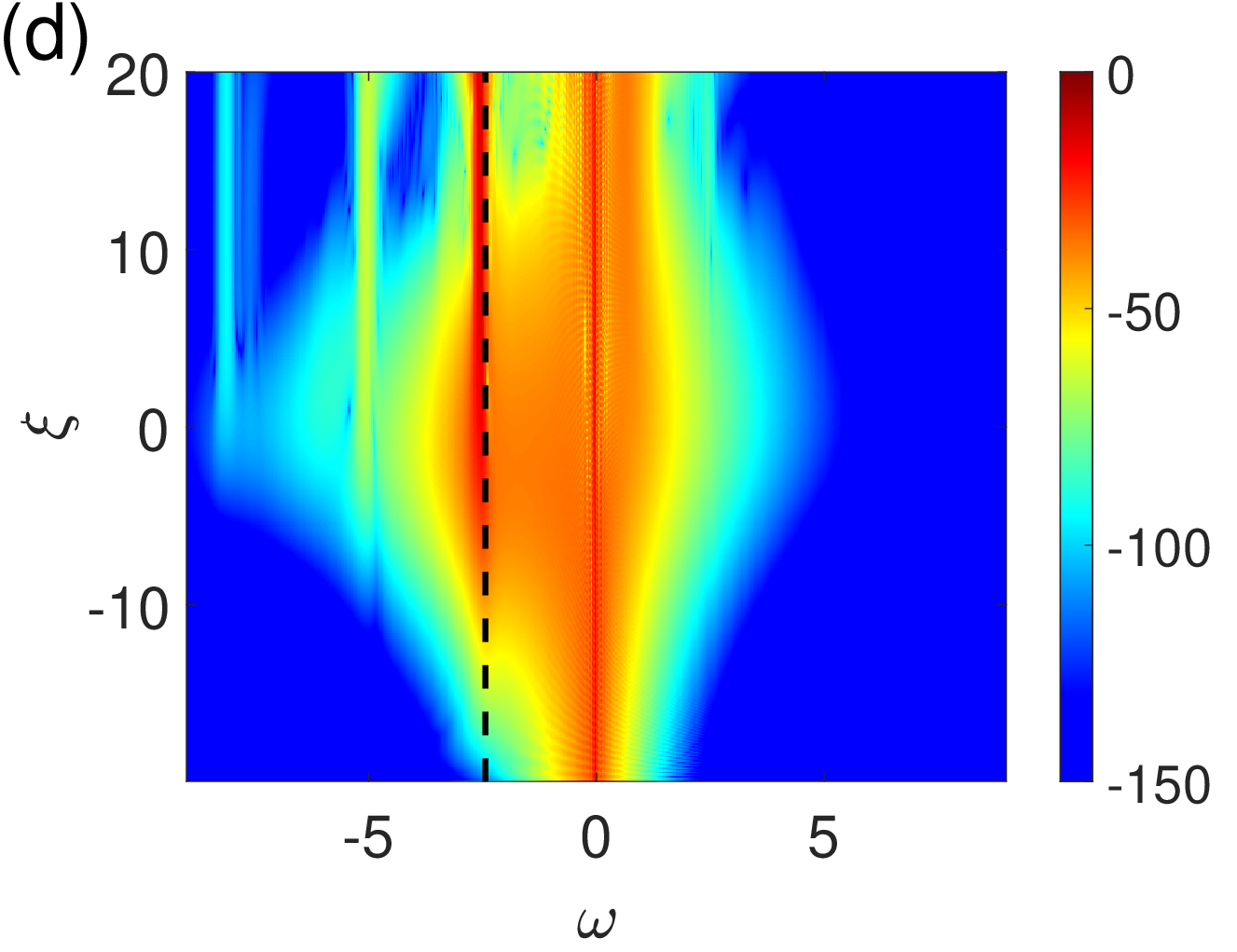}
    \caption{Pseudo-color plot of the spatio-temporal evolution of intensities (a) at FF $|u_1|^2$ and (b) at SH $|u_2|^2$ components of a typical walking Peregrine soliton in the $(\tau,\xi)$ plane. (c,d): corresponding evolution of the FF (c) and SH (d) Fourier spectra (in log scale); the dashed black line marks the predicted resonant frequencies $\omega_{1,FC}^+$ in (c) and $\omega_{2,RR}^+$ in (d). Here $\beta_1=\beta_2=1$, $v=7.5$, $\delta k=-15$.}
    \label{fig:cascading}
\end{figure}

\st{Similarly to the cubic case, such two resonant frequencies are not symmetric around the soliton frequency (they readily become symmetric for stationary solitons, $v_p=0$, implying $v=0$).
Also note that, for small values of the group-velocity mismatch $v$, the constraint $\beta_1\beta_2<0$ holds, which requires GVDs of different signs. This constraint, however, is no longer required as the walk-off $v$ increases. Indeed, for sufficiently large values of $v$, real solutions $\omega_{2,RR}^{\pm}$ do exist for GVDs of the same sign, as shown in the example of Fig. \ref{fig:cascading}.} 

In Fig. \ref{fig:cascading}(a) and (b) we display the evolution of a typical radiating Peregrine soliton in the $(\tau,\xi)$ plane for both the FF and the SH, as obtained by integrating numerically Eqs. \eqref{eq:Quad1} - \eqref{eq:Quad2} with initial value given by the exact solution [Eqs. \eqref{eq:PS1} - \eqref{eq:PS2}] at $z=-30$. In Fig. \ref{fig:cascading}(c) and (d) we show the corresponding Fourier spectral evolutions. In all this example we set $\beta_1=\beta_2=1/2$, $v=7.5$ and $\delta k=-15$ (and hence $\kappa=-1/15$ and $v_p=1/15$). We see significant RR at the SH, emitted at the peak intensity of the Peregrine soliton at $z=0$, preferentially towards positive times. The corresponding frequency detuning, highlighted in Fig. \ref{fig:cascading}(d), is $\omega_{2,RR}^+=-2.4$. The other root, $\omega_{2,RR}^- = -12.4$ is too far detuned to have a noticeable effect on the spectrum.

Once the RR is emitted around the SH, a corresponding radiation arounds the FF can be generated via frequency conversion processes of non-degenerate downconversion type, which can be described as follows. By conveniently introducing the real-world detunings $\Omega_{1,2}=t_0^{-1} \omega_{1,2}$, the RR detuned from the SH by $\Omega_{2,RR}^\pm$ corresponds to photons at physical pulsations $2\omega_0 + \Omega_{2,RR}^\pm$. Such photons can mix through three-photon difference frequency conversion processes $(2\omega_0+\Omega_{2,RR}^\pm)-\omega_0=\omega_0 + \Omega_{1,FC}^\pm$. The conservation of energy, implies equal detunings from the FF and the SH, that is, $\Omega_{1,FC}^\pm = \Omega_{2,RR}^\pm$. 
\st{The RR around the FF, having secondary origin through down-conversion, turns out to be weaker than that around the SH, but is nevertheless quite prominent as shown in Fig. \ref{fig:cascading}(c). In this case, in terms of the normalized frequencies, we find $\omega_{1,FC}^+ = \omega_{2,RR}^+ = -2.4$, which agrees quite well with the RR observed numerically in Fig. \ref{fig:cascading}(c,d). We emphasize that the radiative mechanism turns out to be quite efficient, as the spectrum peaks at the resonant frequency reach values comparable to those of the Peregrine peak, especially in the SH case.}

\begin{figure}
\centering
\includegraphics[width=0.4\textwidth]{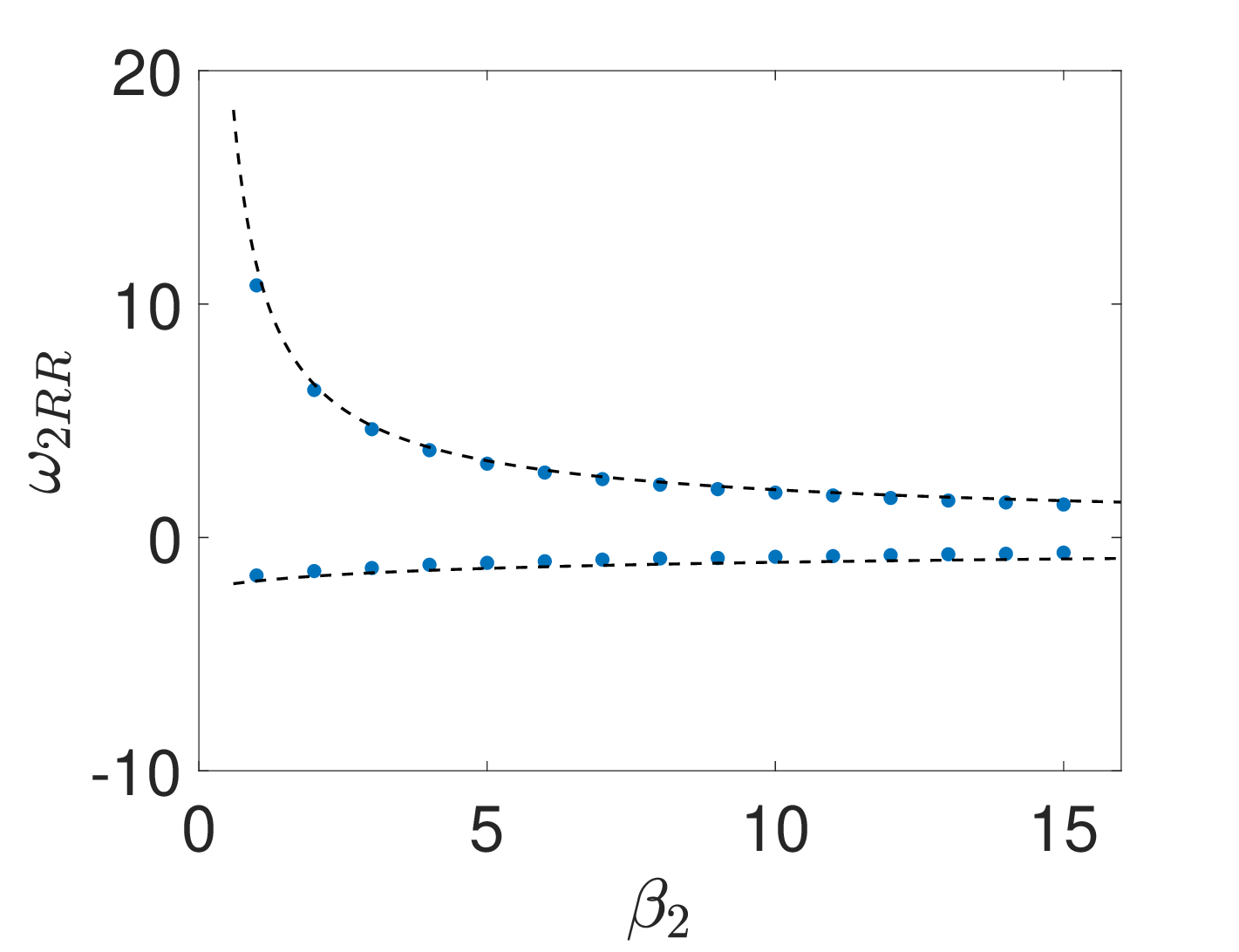}
\caption{RR frequencies $\omega_{2,RR}=\omega_{2,RR}^\pm$ versus GVD at SH $\beta_2$ at $v=5$, comparing theoretical predictions (dashed black curves) and numerical simulations (blue dots). Here $\beta_1=-1$, $\delta k=20$.}
\label{fig:cascRRbeta2} \end{figure}

Additionally, we can study the effect on the resonant frequencies of varying the parameters of the model. Looking at Eq. \eqref{eq:omegawalking}, it only presents real roots if
\begin{equation}\label{eq:CondRealRRbeta2}
    \beta_2\ge -\dfrac{(v_p-v)^2}{2(\delta k+2\kappa + 2\kappa_{loc})} \approx -\dfrac{v^2}{2\delta k}\,.
\end{equation}
In Fig. \ref{fig:cascRRbeta2} we summarize the outcome of numerical simulations in the range of $\beta_2>0$ at $v=5$ (other parameters in the caption). \st{In this case, the dominant (negative) frequency stays nearly constant with $\beta_2$, while the other  frequency, %lower than a certain threshold value, 
turns out to be much more sensitive to $\beta_2$. As the value of $\beta_2$ is reduced below a certain threshold, the RR frequency becomes too large to be noticeably excited since the RR is no longer efficiently seeded by the spectral broadening of the soliton.}

\begin{figure}[h]
\centering
\includegraphics[width=0.4\textwidth]{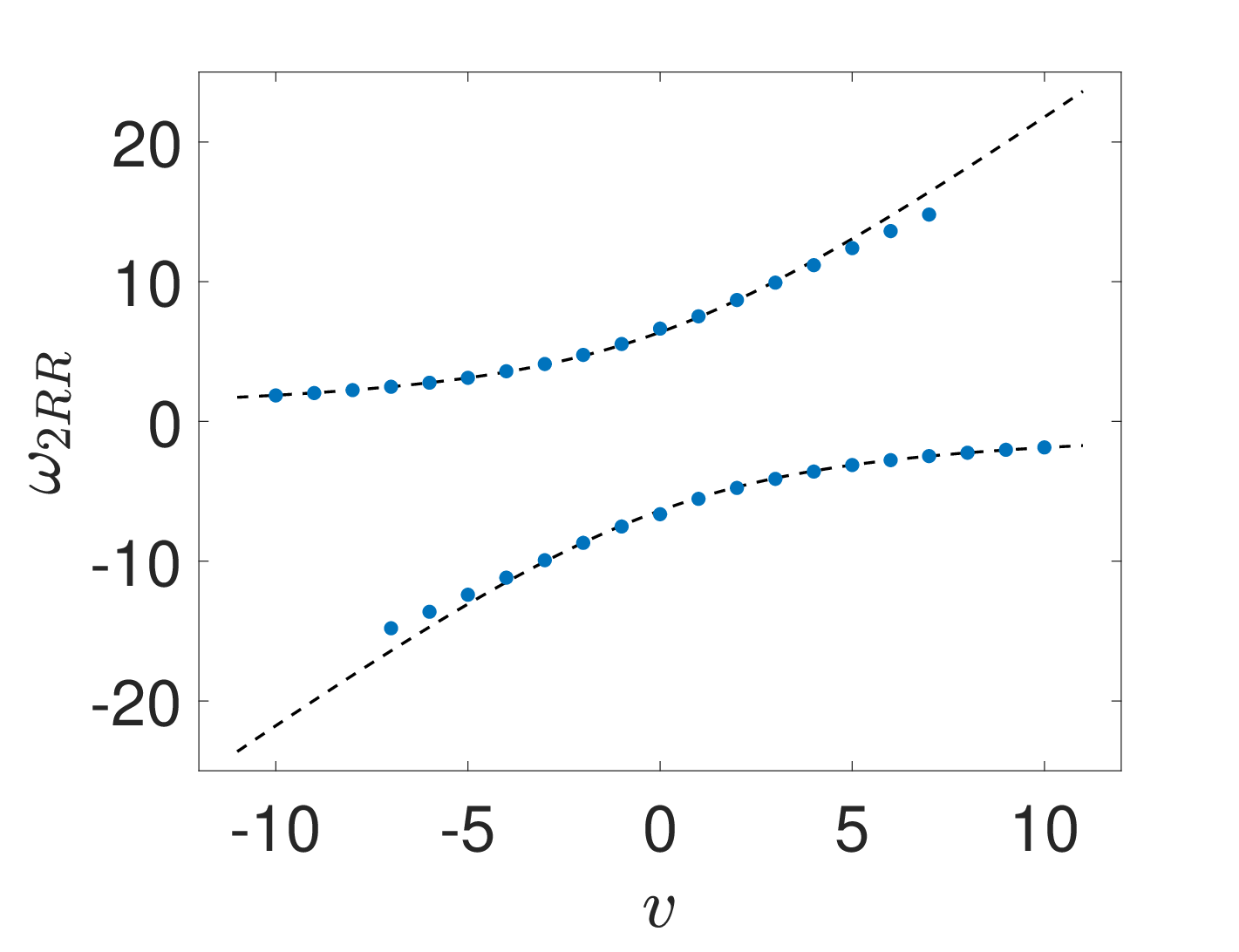}
\caption{RR frequencies versus normalized walk-off $v$, for $\beta_2=1$, comparing theoretical prediction (dashed black curves) with numerical simulations (blue dots). Here $\beta_1=-1$, $\delta k=20$.}
\label{fig:cascRRvpos} \end{figure}

\begin{figure}[h]
\centering
\includegraphics[width=0.4\textwidth]{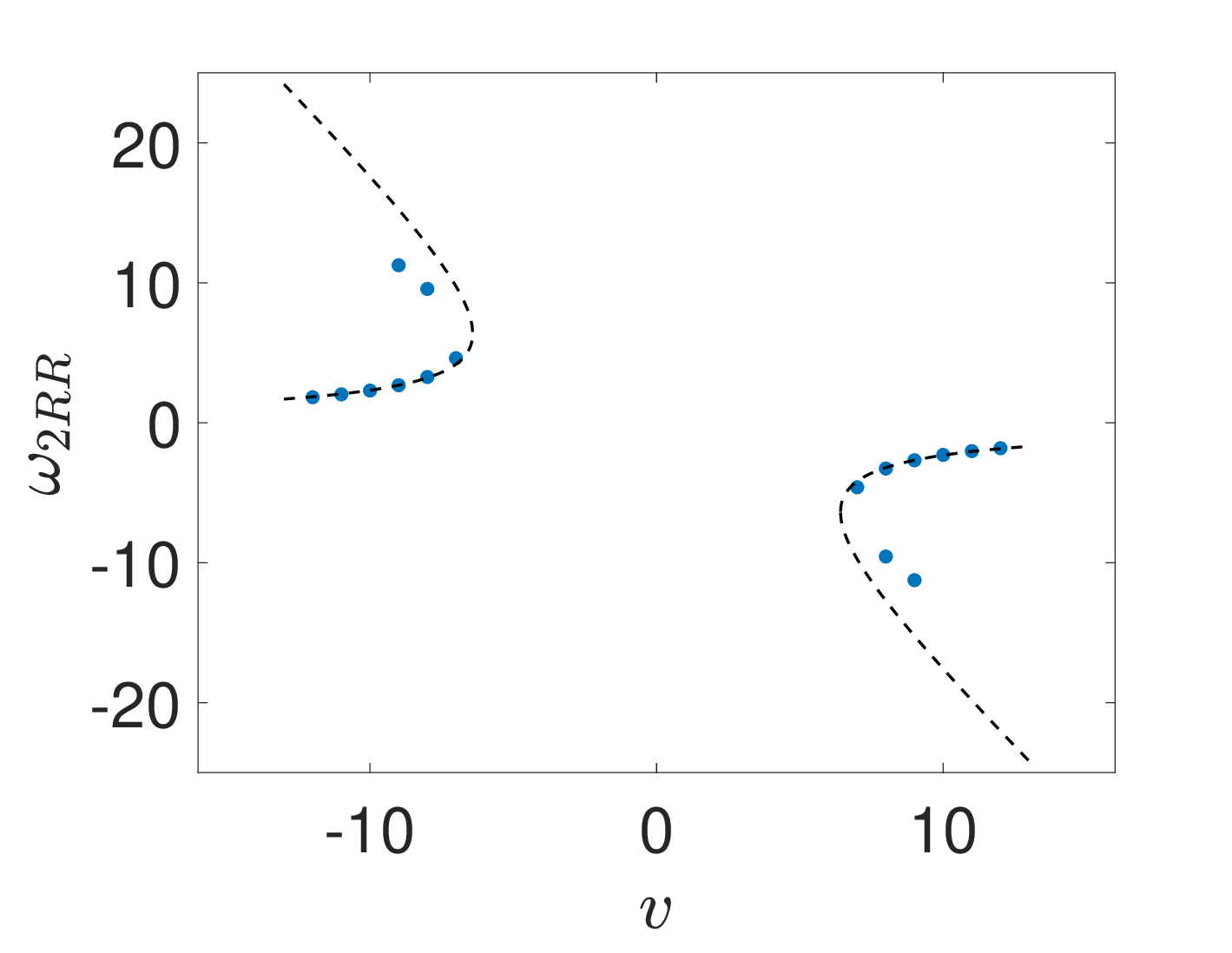}
\caption{As in Fig. \ref{fig:cascRRvpos} for $\beta_2=-1$. Here $\beta_1=-1$, $\delta k=20$.
%Resonant frequencies versus normalized walk-off $v$ for $\beta_2=-1$: theoretical prediction (dashed black curves) are compared with numerical simulations (blue dots). Here $\beta_1=-1$, $\delta k=20$.
}
\label{fig:cascRRvneg} \end{figure}

Finally, we can study how the walk-off affects the RR by varying $v$ under a set of fixed parameters. As shown by Eq. \eqref{eq:CondRealRRbeta2}, for $\beta_2>0$ (normal GVD) there always exist real roots irrespective of the value of $v$. However, when $\beta_2<0$ (anomalous GVD) real roots exist only under the (approximate) condition
\begin{equation}
    |v|> v_{min} \approx \sqrt{2 \delta k |\beta_2|}\,.
\end{equation}
Figure \ref{fig:cascRRvpos} shows the variation of the radiated frequency with $v$ \st{for normal GVD at SH ($\beta_2=1$)}. The frequency pair is situated centrosymmetrically with respect to $v=0$. Numerically, the frequency pair is excited in the range $|v|<7$, beyond which one of the frequencies is too far from the Peregrine soliton to be excited.

Figure \ref{fig:cascRRvneg} shows the variation for the same parameters except \st{for an anomalous GVD at SH ($\beta_2=-1$)}. As shown a gap in walk-off exists, i.e. $|v|<\sqrt{40}$, where the quadratic Peregrine soliton is not resonant with dispersive waves. Interestingly, in this case both resonant frequencies for the same choice of parameters stand on one side of the pump frequency, both \st{positive} when $v<0$ and both  \st{negative} when $v>0$.

%%%%%%%%%%%%%%%%%%%%%%%%%%%%%%%%%%%%%%%%%%%%%%%%%%%

\subsection{Non-cascading regime}
We will now move on to study Peregrine solitons without resorting to the cascading regime. For that, we will consider an anomalous GVD $\beta_1=-1$ for the FF and a normal GVD $\beta_2>0$ for the SH. In this framework, the resulting Peregrine solitons will be still localized on both time and space, and they will present a peak position with more than twice the amplitude of the background.\cite{Kibler2010,Chen2017,Chen2018} This Peregrine solution of Eqs. \eqref{eq:Quad1} and \eqref{eq:Quad2} can be approximated as\cite{Kibler2010,Chen2017,Bu2024}
\begin{gather}
    u_1(\xi,\tau) = A \left[ 1 - \dfrac{2i\gamma\xi + 1/A^2}{\gamma\theta^2 + \gamma^2 A^2 \xi^2 + 1/(4A^2)} \right] \exp\left[ i(\kappa\xi - \Omega\theta) \right]\,,\\
    u_2(\xi,\tau) = \gamma u_1^2 e^{i\delta k\xi}\,,
\end{gather}
where $A$ is the amplitude of the background, $\theta=\tau-v_p\xi$,
\begin{equation}
    v_p=\dfrac{v\beta_1}{\beta_1-2\beta_2}\,,\quad \kappa=\gamma A^2 - \dfrac{v_p^2}{2\beta_1}\,,\quad \Omega=\dfrac{v_p}{\beta_1}\,,
\end{equation}
and
\begin{equation}
    \gamma = \dfrac{vv_p - \beta_1 \delta k - \sqrt{8A^2\beta_1^2 + (vv_p - \beta_1 \delta k)^2}}{4 A^2\beta_1}\,.
\end{equation}
Proceeding as in the cascading limit, we introduce plane waves of the form $\exp(i\kappa_{\text{lin},\{1,2\}}\xi-i\omega_{1,2}\tau_p)$ and equate their wavenumber to that of the nonlinear waves, $\kappa_{\text{lin},1}=\kappa+\kappa_{loc}$, $\kappa_{\text{lin},2}=\delta k + 2\kappa + 2\kappa_{loc}$ (also known as the phase-matching condition for the RR),\cite{Akhmediev1995} where the local correction $\kappa_{loc}$ reads
\begin{equation}
    \kappa_{loc} = \dfrac{8A^2\gamma}{3}\,.
\end{equation}
The resulting conditions read
\begin{gather}
    \gamma A^2 - \dfrac{v_p^2}{2\beta_1} = \dfrac{\beta_1 \omega_1^2}{2} - \omega_1 v_p\,,\\
    \delta k + 2\gamma A^2 - \dfrac{v_p^2}{\beta_1} = \dfrac{\beta_2\omega_2^2}{2} + \omega_2 (v-v_p)\,.
\end{gather}

\colblue{Also similarly to the previous case, the condition for the FF component does not yield real roots whereas frequencies must be real quantities. Therefore the RR will be generated via frequency down-conversion from the RR driven by the SH, which admits the real roots}
\begin{equation}
    \omega_2^\pm = \dfrac{2v_p}{\beta_1} \pm \sqrt{\dfrac{2(2A^2\beta_1\gamma - vv_p + \beta_1 \delta k)}{\beta_1\beta_2}}\,.
\end{equation}
Applying a process analogous to the one employed in the cascading limit, the down-converted RR around the FF is found to occur at frequencies
\begin{equation}
    \omega_1^+ = \dfrac{3\omega_2^+ - \omega_2^-}{4}\,,\quad \omega_1^- = \dfrac{3\omega_2^- - \omega_2^+}{4}\,.
\end{equation}
Note that both for the FF and the SH the resonant frequencies are located symetrically with respect to the central frequencies $\Omega=v_p/\beta_1$ and $2\Omega = 2v_p/\beta_1$, respectively.

%-------------------- Fig. 8
\begin{figure}[h]
    \centering
    \includegraphics[width=0.35\textwidth]{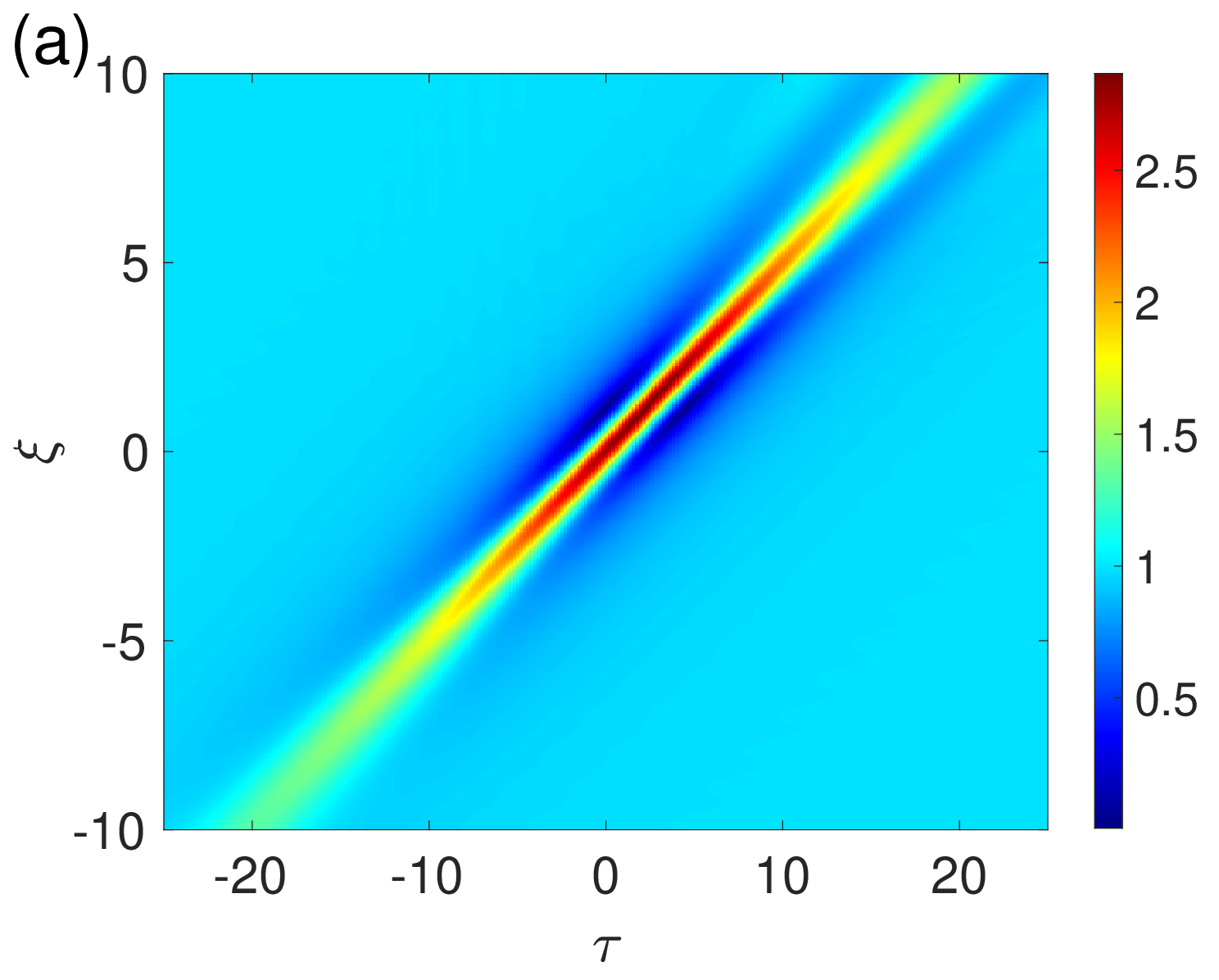}
    \includegraphics[width=0.35\textwidth]{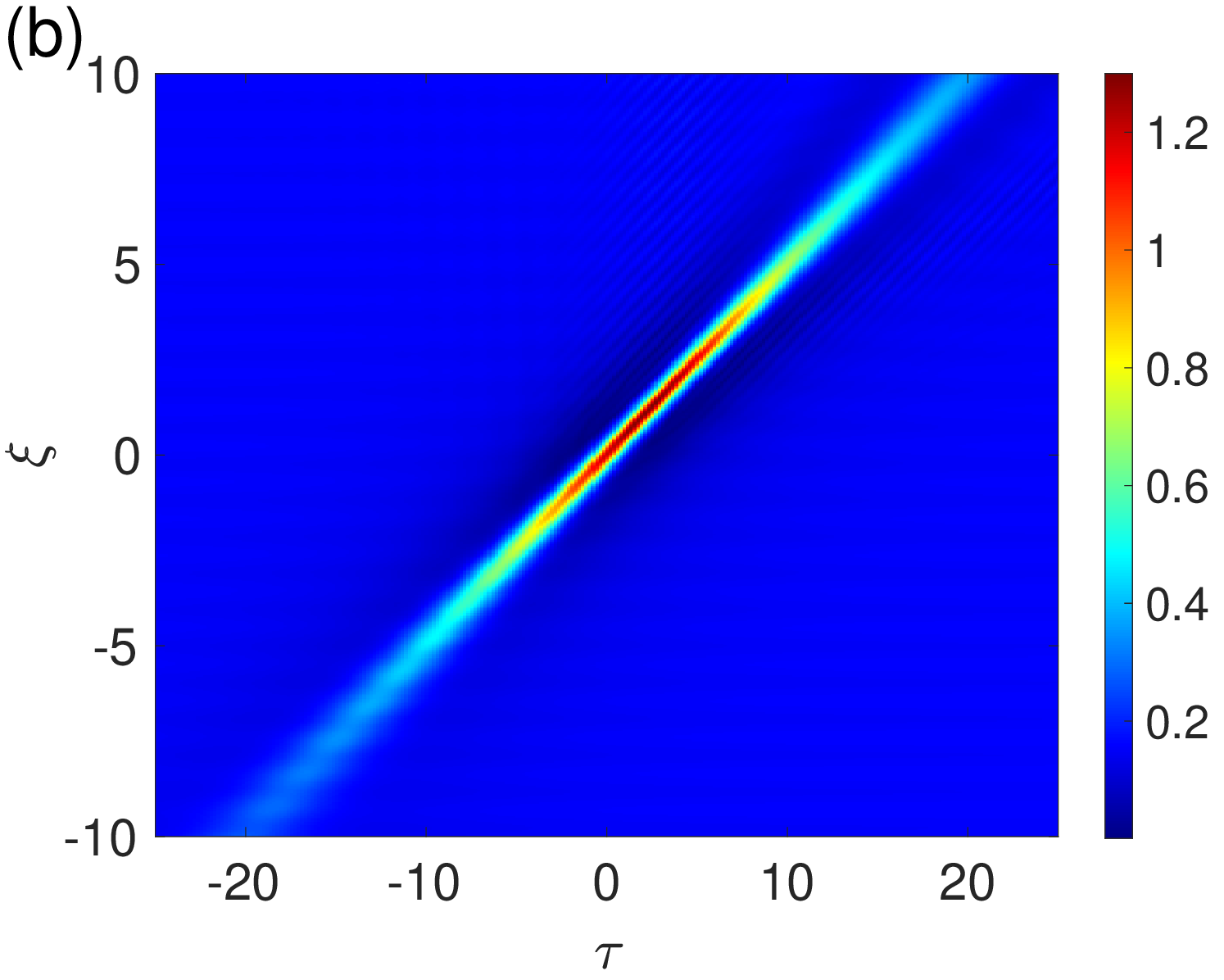}
    \includegraphics[width=0.35\textwidth]{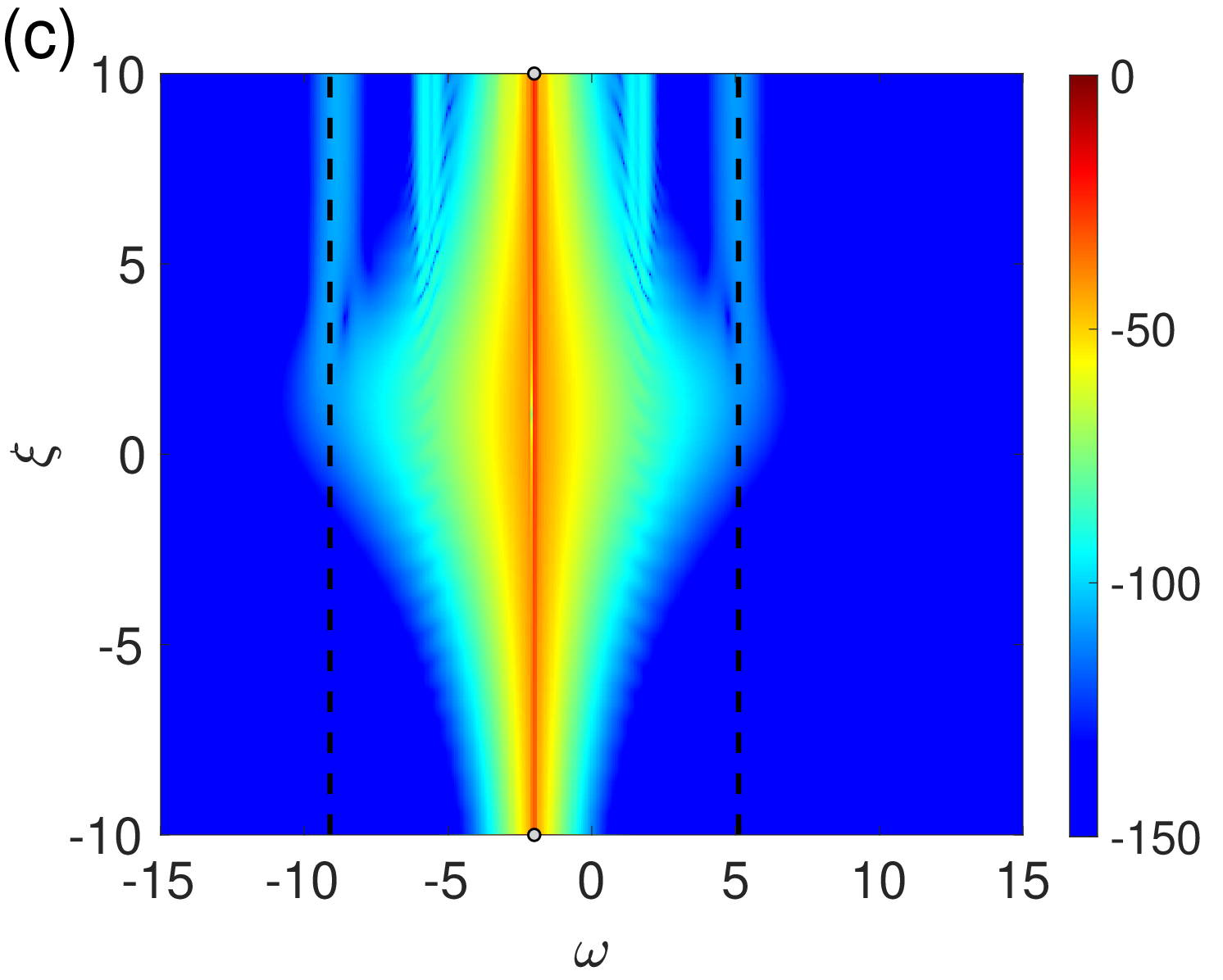}
    \includegraphics[width=0.35\textwidth]{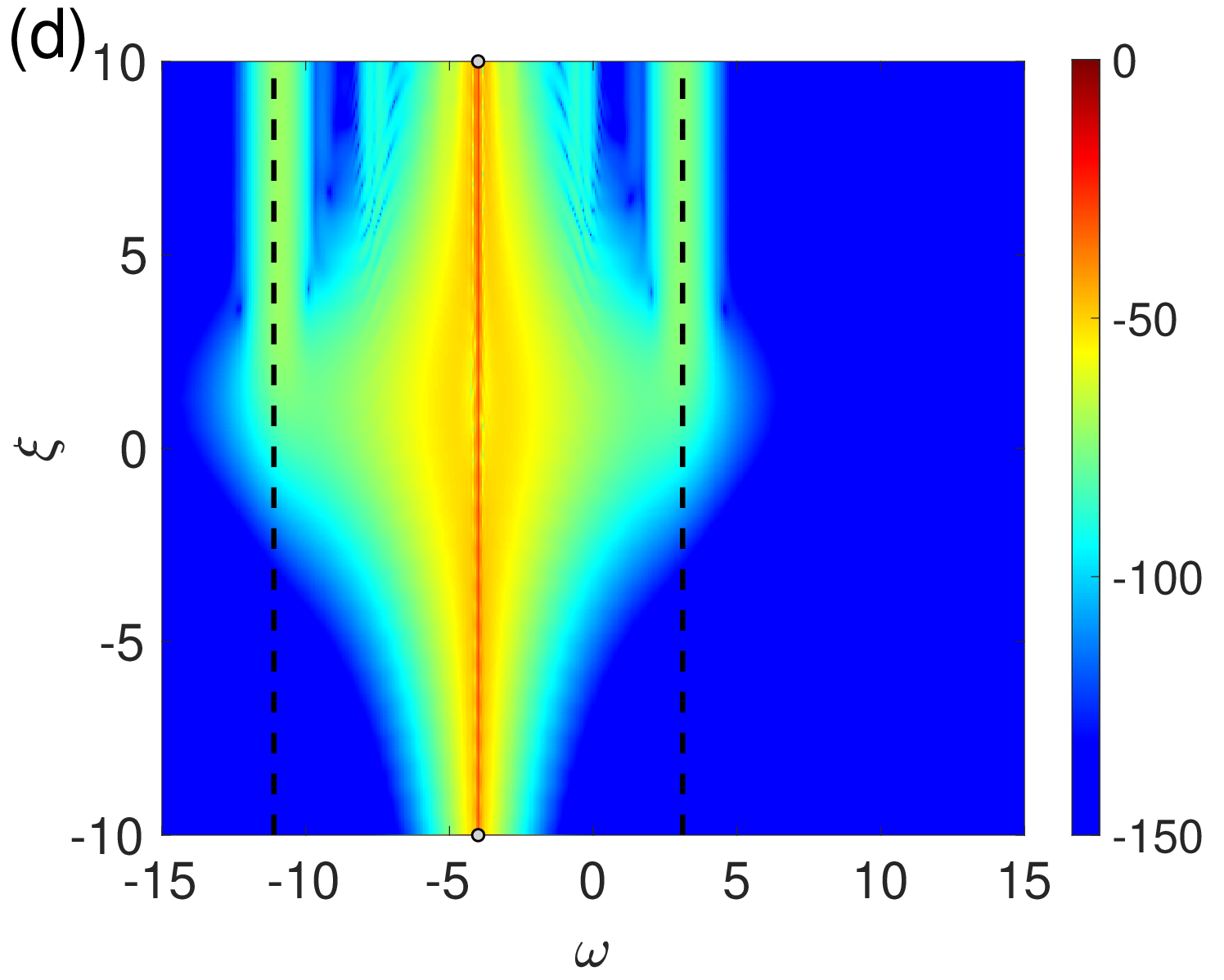}
    \caption{Optical RR emitted by the PS under phase-matching $\delta k=0$, with a large temporal walk-off $v=3$, $\beta_1=-1$, $\beta_2=1/4$ and $A=1$. 
    \st{(a,b): Temporal evolutions of intensities at FF, SH, respectively;  (c,d): spectral (log scale) evolutions at FF, SH, respectively, compared with predictions (dashed black vertical lines) from Eqs. (24-25).}
    }
\label{fig:noncascading1} \end{figure}

In Fig. \ref{fig:noncascading1} we show the RR of a Peregrine soliton under the phase-matched condition $\delta k=0$, with a large temporal walk-off $v=3$. The analytic solutions presented above are consistent with the numerical results, even in the central peak, which exhibits the higher compression. Strong RR appear after the central peak $\xi=0$, both in the SH component and later on in the FF component via downconversion. Its frequency is consistent with our predictions, which for these values of the parameters translate into $\omega_2^+=3.11$ and $\omega_2^-=-11.11$ for the SH, and $\omega_1^+ = 5.11$ and $\omega_1^-=-9.11$ for the FF, which are highlighted by the dashed vertical black lines in Fig. \ref{fig:noncascading1}(c) and (d).

This strong RR may be related to the transient property of the Peregrine solitons, which undergoes and intense compression at the peak to then radiate as it recovers from it. It is worth noting that more resonant frequencies that are not predicted by our computations appear in Fig. \ref{fig:noncascading1}(c) and (d), which require further research.

%%%%%%%%%%%%%%%%%%%%%%%%%%%%%%%%%%%%%%%%%%%%%%%%%%%%%%

\section{Conclusions} \label{conclu}
\redc{To summarize, we have provided}  our recent advances in \st{understanding} the RR emitted by Peregrine solitons \redc{in either cubic or quadratic media, with extra new numerical simulations}. 

In the case of cubic media, we have shown that Peregrine solitons can emit linear dispersive waves at certain resonant frequencies in presence of higher-order dispersion, \redc{similarly to the case with classical bright and dark solitons}. However, unlike the \redc{latter}  soliton case, the radiated frequency can be effectively affected by the local contribution to the intrinsic wavenumber of the Peregrine soliton. This contribution is not unique to the Peregrine \redc{soliton}, instead it is shared by other kinds of rogue-type solutions, in particular, \redc{Akhmediev breathers} and \redc{Kuznetsov}-Ma solitons, possibly describing other rogue wave dynamics.

In the case of quadratic media, our results \redc{offer} a wider perspective of the phenomenon of \redc{soliton} RR, showing that \redc{it can be driven by the GVD of the SH component and then occurs simultaneously on the FF component as well via} concomitant difference-frequency generation processes.  \redc{Our analytical criterion obtained from the phase-matching condition can accurately predict the resonant frequencies at which the linear dispersive waves are emitted}. \redc{Different from} other RR mechanisms, typically driven by higher-order dispersions, the present mechanism can also be employed for the case of spatial \redc{
solitary} waves, which are described by \redc{coupled} second-harmonic generation equations \redc{but with anomalous GVD replaced by the spatial diffraction}. Thus, the introduced mechanism can lead to a rare case of radiating spatial solitons, as well as induce reshaping of the spectrum of spatio-temporal bullets in bulk media.

\section*{Acknowledgments}
This work was supported by the Progetti di Ricerca di Interesse Nazionale (PRIN) (Project No. 2020X4T57A), the PRIN funded by European Union--Next Generation EU (Project 20222NCTCY), the National Natural Science Foundation of China (Grants No. 12374301 and No. 11974075) and the China Scholarship Council (CSC202006090086).

%%%%%%%%%%%%%%%%%%%%%%%%%%%%%%%%%%%%%%%%%%%%%%%%%%%%%%%%%%%%%%%%%%%%%%

\bibliographystyle{ieeetr}

\bibliography{rrreview}

\end{document}